\documentclass[review]{elsarticle}

\usepackage{geometry}
\usepackage{array}
\usepackage{amsmath}
\usepackage{amsmath}
\usepackage{amssymb}
\usepackage{amsmath,amssymb}
\usepackage[amsmath,thmmarks,hyperref]{ntheorem}
\usepackage{mathrsfs}
\usepackage{extarrows}
\allowdisplaybreaks[4]
\usepackage{abstract}
\usepackage{indentfirst}
\usepackage[thmmarks,amsmath]{ntheorem}
\usepackage{array}
\usepackage{supertabular}
\usepackage{amstext}
\usepackage{setspace}
\usepackage{graphicx}
\usepackage{epstopdf}
\usepackage{colortbl}
\usepackage[utf8]{inputenc}
\usepackage[T1]{fontenc}
\usepackage[noend]{algpseudocode}
\usepackage{algorithmicx,algorithm}

\usepackage{multirow}
\usepackage{caption}
\usepackage{booktabs}
\usepackage{float}
\usepackage{tikz}
\newcommand*{\circled}[1]{\lower.7ex\hbox{\tikz\draw (0pt, 0pt)%
    circle (.5em) node {\makebox[1em][c]{\small #1}};}}

\usepackage[labelsep=quad,justification=centerlast,labelfont=footnotesize,textfont=footnotesize]{caption}
\usepackage{subfigure}

\journal{Journal of \LaTeX\ Templates}

\begin{document}
\large
\numberwithin{equation}{section}
\newtheorem{theorem}{Theorem}[section]
\newtheorem{lemma}[theorem]{Lemma}
\newtheorem{Remark}[theorem]{Remark}
\newtheorem{assumption}[theorem]{Assumption}
\theoremstyle{definition}
\newtheorem{mydefinition}{Definition}
\theoremstyle{nonumberplain}
\theoremheaderfont{\bfseries}
\theorembodyfont{\normalfont}
\newtheorem{Proof}{Proof.}

\title{ The Numerical Simulation of Quanto Option Prices Using Bayesian Statistical Methods}

\author[ouraddress,liaddress]{Lisha Lin}
\author[liaddress]{Yaqiong Li \corref{mycorrespondingauthor}}
\cortext[mycorrespondingauthor]{Corresponding author}
\ead{yqli@hnu.edu.cn}
\author[ouraddress,liaddress]{Rui Gao}
\author[wuaddress]{Jianhong Wu}
\address[ouraddress]{College of Mathematics and Econometrics,
Hunan University, Changsha, 410082, China}
\address[liaddress]{College of Finance and Statistics,
Hunan University, Changsha, 410079, China}
\address[wuaddress]{Department of Mathematics and Statistics,
York University, Toronto, ON, M3J 1P3, Canada}

\maketitle

\begin{abstract}
\large
In the paper, the pricing of Quanto options is studied, where the underlying foreign asset and the exchange rate are correlated with each other. Firstly, we adopt Bayesian methods to estimate unknown parameters entering the pricing formula of Quanto options, including the volatility of stock, the volatility of exchange rate and the correlation. Secondly, we compute and predict prices of different four types of Quanto options based on Bayesian posterior prediction techniques and Monte Carlo methods. Finally, we provide numerical simulations to demonstrate the advantage of Bayesian method used in this paper comparing with some other existing methods. This paper is a new application of the Bayesian methods in the pricing of multi-asset options.


\end{abstract}




\section{\large Introduction}
With the rapid and deep development of globalization, the Quanto option has received much attention from both investors and financial institutions, since it provides a platform for domestic investors to manage multinational risks and to obtain exposures of foreign assets, and its price depends on both the price changes of foreign asset and the fluctuations of exchange rate.

Quanto option is one kind of multi-asset exotic options, whose payoff is converted into another currency as the underlying asset is traded. 
Some recent studies have extended the underlying models of Quanto options beyond multivariate geometric Brownian motions studied in previous literatures \cite{Reiner(1992),Dravid(1993),Dai(2004),Yaqiong(2011)}. These extensions are needed due to the limitations of Black-Scholes \cite{BlackandScholes(1973)} quanto model pointed out in  \cite{Long(2015),Giese(2012),Fallahgoul(2018),Kim(2015)}, including the phenomena of jumps, heavy tails and skewness etc..
Teng et\,al.\,\cite{Long(2015)} derived a closed-form formula, calibration and hedging strategy for Quanto options under a dynamic correlation assumption by employing a dynamic correlation model. 
Giese \cite{Giese(2012)} obtained explicit solutions for the prices of Quanto options under a stochastic volatility model. Kim et al.\,\cite{Kim(2015)} studied Quanto option pricing using a multivariate normal tempered stable process that characterizes fat-tailedness and asymmetric dependence between underlying asset returns and exchange rate returns. 
In comparison, Fallahgoul et al.\,\cite{Fallahgoul(2018)} developed a multivariate L\'evy model to capture more features, including jumps, skewness and fat-tailedness, observed in real markets for stock prices and exchange rates. This model in \cite{Fallahgoul(2018)} shows superiority to the normal tempered stable process \cite{Kim(2015)} in terms of fitting market distribution and pricing Quanto options. However, while the above-mentioned complicated model structures for Quanto options are useful to capture the real market situation, they also lead to complications in the model estimation problems. 
\par
Insteading of extending the Quanto option pricing into a more general setting, here we focus on explaining how to compute Quanto option prices via Bayesian statistical inference, where the processes of underlying asset price and exchange rate are described by two correlated geometric Brownian motions. Therefore, this paper serves as a basis for studying the more complicated Quanto option pricing problems in a Bayesian framework. In addition to the volatility parameters of the underlying asset and the exchange rate, the correlation between both is also estimated using Bayesian method in this paper. As illustrated in Refs.\,\cite{Long(2015),Dimitroff(2012)}, the increase of the number of underlying assets included in options is accompanied by the incorporation of the correlation between assets, which requires special consideration and involves the simulation of a system of correlated SDE models in numerical experiments. 
Dimitroff et al.\,\cite{Dimitroff(2012)} provided detailed computation methods of the correlation of asset-asset, asset-volatility and volatility-volatility by using the historical data of underlying assets under a multi-asset Heston model. Giese \cite{Giese(2012)} compared the historical correlation calculated from the asset information with the method constructed in \cite{Dimitroff(2012)}. Vasiliki et al.\,\cite{Vasiliki(2005)} proposed a method about how to predict implied correlation from option prices. However, the estimation methods for the correlation parameter mentioned above are considered in frequentist approaches. Bayesian methods treat the unknown parameters as random variables and offers a reasonable way to account for parameter uncertainty. This paper develops a fully detailed Bayesian approach for the estimation of parameters entering the underlying asset dynamics and the pricing of four different types of Quanto options.

Once the diffusion model of the price processes of underlying assets is defined, the estimation of unknown parameters entering those processes plays a central role for further pricing options. Traditionally, the method of maximum likelihood estimation (MLE) is employed, but it always results in biased estimated option prices except for at-the-money options as illustrated in \cite{Knight(1997)}. A limitation of MLE method is that it only provides a point estimation of parameters. Also, for many diffusion processes, whose transition densities are rather complicated or not given in a closed form, the MLE method is infeasible.
More recently, the Bayesian method, which allows for the flexibility of prior information on parameters, has become a mainstream in accounting for parameter uncertainty \cite{Tunaru(2017),Kattwinkel(2017)}. 
This paper provides an alternative way to study multi-asset option pricing and it is expected to reduce the pricing error by adopting Bayesian methods. 
\par
Some Bayesian analyses in an option pricing framework have been conducted.
Jacquier and Jarrow \cite{Jacquier(2000)} took parameter uncertainty and model error into account to perform the Bayesian estimation of contingent claim models. Their method was implemented in the case of basic and extended B-S models, and adoption of this method in empirical applications showed that the extended B-S model exhibits an improvement in pricing bias compared to the basic B-S model in the in-sample case. However, this result is not valid anymore in the out-of-sample case.
Martin et al.\,\cite{Martin(2010)} conducted posterior inference for a range of returns models and obtained estimations from option data. Their inference incorporated both the parameter and model uncertainties. Rombouts and Stentoft \cite{Rombouts(2014)} computed European call option prices by aggregating the predicted density of underlying asset returns until maturity, where the underlying return process is governed by an asymmetric heteroskedastic normal mixture model. They applied the method to the S$\&$P500 index, and compared the Bayesian inference with classical inference for the specific two components mixture model in terms of parameter estimation and option pricing performances. The numerical experiments provided evidence that there might be potential advantages for using Bayesian inference when less return data is available. Gao et al.\,\cite{Gao(2019)} performed Bayesian statistical inference for the pricing of European call option with stock liquidity in an incomplete market. Their numerical experiments with applications to the S$\&$P500 index option indicated the potential advantages of Bayesian methods compared with traditional statistical methods in parameter estimations as well as option pricing. 
\par
Closely associated with our paper are the studies by Karolyi \cite{Karolyi(1993)} and Darsino and Satchell \cite{TheoDarsinoandStephenSatchell(2007)}. Both devoted to performing Bayesian inference for the pricing of European call option. In \cite{Karolyi(1993)}, Karolyi viewed the cross-sectional group of stock return volatilities as a source of prior information, and derived the posterior density of volatility that improves the estimation precision of option price. Furthermore, Darsino and Satchell in \cite{TheoDarsinoandStephenSatchell(2007)} involved the randomness arising from both the underlying asset price process and the volatility to deduce the prior and posterior densities for option prices. They, later in \cite{Darsinos(2007)}, established a Bayesian predictive framework for B-S option prices, which allows for the collection of information from historical return data and from implied volatility of reported option prices in a rigorous way.
\par
Despite its popularity and relative superiority reported in literatures \cite{LC(2017),Wei(2017)} in terms of pricing options on one asset, Bayesian method has not been extensively used for the pricing of multi-asset options. This paper aims to extend the Bayesian inference framework from the pricing of one-asset options in \cite{Karolyi(1993),TheoDarsinoandStephenSatchell(2007)} to two-asset options. We perform Bayesian statistical inference on the pricing of Quanto options, which have two underlying assets, with the randomness arising from volatilities of foreign asset and exchange rate and the correlation between them. The increase of the number of underlying asset inevitably increases the number of model parameters, and thus potentially makes the Bayesian inference more complicated than that of the one-asset options \cite{Karolyi(1993),TheoDarsinoandStephenSatchell(2007)}. Especially, the correlation arising from the increase of underlying asset is also treated as a random variable and estimated within the Bayesian framework in a natural way.

Quanto options are of great significance for the increasingly popularity of global asset investments, which enable investors from different countries to settle payoffs in their own currencies. We take four different types of European Quanto options as examples to illustrate how to compute option prices by using the predictive density being a by-product of Bayesian inference. 
Given that the posterior density is generally not in forms that we are familiar with and directly sampling from it is infeasible, developing an efficient method used for posterior simulation is an important issue.  
Markov Chain Monte Carlo (MCMC) methods are a class of commonly used algorithms that can be used to simulate from unknown distributions by producing a Markov chain that converges to the desired posterior distribution. The Gibbs sampling \cite{Geman(1984)} and the Metropolis-Hastings algorithms \cite{Chib(1995)} are two widely used MCMC approaches, where the Gibbs sampling algorithm \cite{Geman(1984)} draws samples sequentially from the full conditional posterior distributions, and the Metropolis-Hastings algorithm \cite{Chib(1995)} simulates samples from a carefully chosen proposal distribution when the posterior density is not a standard form, and then accepts or rejects theses samples according to the acceptance probability. The combination of the Gibbs sampling algorithm and the Metropolis-Hastings algorithm, i.e.\,the Metropolis with-in Gibbs algorithm \cite{Besag(1995)}, has been proved to be an efficient way for sampling from unrecognizable posterior conditional densities.     
Considering the fact of the multi-dimensional posterior density, this paper adopts the Metropolis with-in Gibbs algorithm \cite{Besag(1995)} with different candidate densities to solve the posterior simulation problem. It is expected that, compared to some existing statistical inference methods, the Bayesian estimation of parameters and option prices developed in the paper have advantages in some respects.
\par
The structure of this paper is as follows. In section 2, we give the basic dynamic models for the price processes of foreign underlying asset and exchange rate. In section 3, we perform Bayesian inference on parameters entering the two dynamic models. In section 4, we present Bayesian prediction methods to compute option prices. In section 5, we report the numerical posterior simulations, and we summarize our conclusions in section 6.

\section{\large Dynamic models for the processes of underlying asset price and exchange rate}
Suppose that $(\Omega, \mathcal{F}, \{\mathcal{F}_t\}_{t\geq 0}, \mathrm{\bf{P}})$ is a filtered probability space with the filtration $\{\mathcal{F}_t\}_{t\geq 0}$ satisfying usual conditions. 
Let $X_t$ and $H_t$ be the foreign asset and the exchange rate (price of foreign currency in domestic currency) defined on the probability space $(\Omega, \mathcal{F}, \{\mathcal{F}_t\}_{t\geq 0}, \mathrm{\bf{P}})$, whose price processes satisfy the following two-dimensional stochastic differential equations
\begin{equation}\label{basic}
\left\{
\begin{aligned}
\mathrm{d}X_t&=\mu_x {X_t}\,\mathrm{d}t+\sigma_x {X_t}\,\mathrm{d}W_x^P(t),
\\ \mathrm{d}H_t&=\mu_h {H_t}\,\mathrm{d}t+\sigma_h {H_t}\,\mathrm{d}W_h^P(t),
\end{aligned}\right.
\end{equation}
where $\mu_x,\,\sigma_x$ are the drift and volatility of $X_t$, $\mu_h,\,\sigma_h$ are the drift and volatility of $H_t$, $W_x^P(t)$ and $W_h^P(t)$ are two correlated Wiener processes with correlation coefficient $\rho$, i.e. $\mathrm{d}W_x^P(t)\mathrm{d}W_h^P(t)=\rho\,\mathrm{d}t$. Let $r_d$ and $r_f$ be the domestic and foreign risk-free interest rate, respectively.
Let $W_h^P(t)\triangleq\rho W_x^P(t) + \sqrt{1-\rho^2} W_y^P(t)$, where $W_y^P(t)$ is a Wiener process independent of $W_x^P(t)$. By Ito$'$s formula, we have solutions $\{X_t\}_{t \geq 0}$ and $\{H_t\}_{t \geq 0}$ for equation (\ref{basic})
\begin{equation*}
\begin{cases}
\setlength\abovedisplayskip{0.5pt}
\setlength\belowdisplayskip{0.5pt}
X_t =X_0 \exp \left\{\left(\mu_x-\displaystyle\frac{1}{2}{\sigma_x}^2\right)t + \sigma_x W_x^P(t)  \right\},
\\ H_t = H_0 \exp \left\{\left(\mu_h-\displaystyle\frac{1}{2}{\sigma_h}^2\right)t + \sigma_h \left(\rho W_x^P(t) + \sqrt{1-\rho^2} W_y^P(t) \right)\right\}.
\end{cases}
\end{equation*}

As pointed out by many researchers \cite{Long(2015),Giese(2012),Fallahgoul(2018),Kim(2015)}, the constant volatility assumption can not characterize the real market situation. Bayesian method considered in the paper will treat the unknown parameters as random variables to account for their uncertainties. In this paper, we consider the randomness from parameters and study the pricing of four different types of Quanto options in a Bayesian framework. The pricing problem of various Quanto options has been discussed by Reiner \cite{Reiner(1992)} \,under the model (\ref{basic}):

$\circled{1}$ Foreign stock call Quanto options struck in domestic currency with payoff function $F^{(1)}$:
{\setlength\abovedisplayskip{1pt plus 3pt minus 7pt} 
\setlength\belowdisplayskip{1pt plus 3pt minus 7pt}
$$F^{(1)}=\max(H_TX_T-K_d,0),$$}
where $H_T$ and $X_T$ are the exchange rate and underlying asset price at maturity $T$, $K_d$ is the strike price in domestic currency. 

$\circled{2}$ Floating exchange rate foreign stock call Quanto options struck in foreign currency with payoff function $F^{(2)}$:
{\setlength\abovedisplayskip{1pt plus 3pt minus 7pt} 
\setlength\belowdisplayskip{1pt plus 3pt minus 7pt}
$$F^{(2)}=H_T\max(X_T-K_f,0),$$}
where $K_f$ is the strike price in foreign currency. 

$\circled{3}$ 
Fixed exchange rate foreign stock call Quanto options struck in foreign currency with payoff function $F^{(3)}$:
{\setlength\abovedisplayskip{1pt plus 3pt minus 7pt} 
\setlength\belowdisplayskip{1pt plus 3pt minus 7pt}
$$F^{(3)}=H_{fix}\max(X_T-K_f,0),$$}
where $H_{fix}$ is some fixed exchange rate determined in advance. 

$\circled{4}$ 
Stock-linked foreign exchange rate call Quanto options with payoff function $F^{(4)}$:
{\setlength\abovedisplayskip{1pt plus 3pt minus 7pt} 
\setlength\belowdisplayskip{1pt plus 3pt minus 7pt}
$$F^{(4)}=X_T\max(H_T-K_H,0),$$}
where $K_H$ is the strike price for exchange rate.

Let  $V^{(i)}$ be the Quanto option price with payoff function $F^{(i)}$, $i=1,2,3,4$. From Ref.\,\cite{Reiner(1992)}, we know that  except for case $\circled{2}$, the analytical formulas of $V^{(1)}, V^{(3)}$ and $V^{(4)} $ are functions of parameters $\sigma_x, \sigma_h, \rho$, while $V^{(2)}$ only depends on $\sigma_x$. Therefore, despite with the underlying models described by Eqs.\,(\ref{basic}), Quanto option prices are conditional on $\sigma_x$, $\sigma_h$, $\rho$ unknown, and how to estimate these parameters is a basic question with the aim of pricing options further. This paper particularly uses Bayesian approach for the statistical inference of unknown parameters and for the predictive price of Quanto option by embedding the uncertainty of parameters in both the posterior distributions and the predictive densities. 
%
%
\section{\large \bf{Bayesian inference on the processes of asset price and exchange rate}}
Given the underlying assets models (\ref{basic}), it is crucial to develop an efficient statistical inference framework to estimate parameters of the basic models, which further act as inputs of pricing options.
We denote by
\begin{align}\label{return}
x_t\triangleq\log\displaystyle\frac{X_t}{X_{t-1}} \quad \mathrm{and} \quad h_t\triangleq\log\displaystyle\frac{H_t}{H_{t-1}}
\end{align}
the individual log return of $X_t$ and $H_t$ between consecutive time interval, for $t=1,2,\cdots,T$. Then the joint probability density function of $x_t$ and $h_t$ is
\begin{align}\label{pdf}
\nonumber p \big(x_t,h_t|&\mu_x,\mu_h,\sigma_x,\sigma_h,\rho \big) = \displaystyle\frac{1}{2 \pi\,\sigma_x\sigma_h \sqrt{1\!-\!\rho^2}}
 \exp \Bigg\{-\displaystyle\frac{\left[x_t-\left(\mu_x\!-\!\frac{1}{2}{\sigma_x}^2\right)\right]^2}{2 {\sigma_x}^2(1\!-\!\rho^2)}
\\ &-\displaystyle\frac{\left[h_t-\left(\mu_h\!-\!\frac{1}{2}{\sigma_h}^2\right)\right]^2}{2 {\sigma_h}^2(1\!-\!\rho^2)}
+\displaystyle \frac{\left[x_t-(\mu_x\!-\!\frac{1}{2}{\sigma_x}^2)\right]\left[h_t-(\mu_h\!-\!\frac{1}{2}{\sigma_h}^2)\right]\rho}{\sigma_x \sigma_h (1\!-\!\rho^2)}\Bigg\}.
\end{align}
%

To simplify the Bayesian inference based on a two-dimensional distribution (\ref{pdf}), we separate the joint distribution of $X_t$ and $H_t$ into two one-dimensional distributions and then consider the two parts in sequence in a Bayesian framework, since
$$p\left(X_t,H_t\right)=p(X_t|H_t)p(H_t),$$
where  \quad \qquad \qquad \qquad
$\log H_t\!\thicksim\!N\,\left[\log H_0\!+\!(\mu_h\!-\!\displaystyle\frac{1}{2}\sigma_h^2)\,t,\,\sigma_h^2\,t\right],$
\vskip2.4mm
\noindent \qquad \qquad \qquad \qquad\qquad
$\log X_t|\log H_t\!\thicksim\!N\,\Big[\,\mu_{x|h},\,(1\!-\!\rho^2)\sigma_x^2\,t\,\Big],$
\vskip2.4mm
\noindent  \qquad \qquad  \qquad $\mu_{x|h}\triangleq\log X_0\!+\!(\mu_x\!-\!\displaystyle\frac{1}{2}\sigma_x^2)\,t\!+\!\rho\displaystyle\frac{\sigma_x}{\sigma_h}\,\left[ \log\frac{H_t}{H_0}\!-\!(\mu_h\!-\!\displaystyle\frac{1}{2}\sigma_h^2)\,t\right],$
\vskip1mm 
\noindent with $N[a,b]$ denoting the normal probability density with mean $a$ and variance $b$.
\vskip1mm
\par
Therefore, our Bayesian inference involves two stages: inferring parameters entering the distribution of $H_t$, followed by inferring parameters entering the distribution of $X_t$ conditional on $H_t$. It is natural to adopt posterior results of the former as priors to infer the latter. Bayesian method will produce a complete posterior density of unknown parameters as well as expected prices of Quanto options with different payoff functions.

In the first stage, when only one asset is involved, this issue is related to the pricing of options on one underlying asset. For example, Karolyi \cite{Karolyi(1993)} and Darsino and Satchell \cite{TheoDarsinoandStephenSatchell(2007)} considered the European call option pricing in a Bayesian framework, where conjugate priors are commonly applied. For comparision, we choose noninformative priors in this section to account for the randomness of parameters.
\par Assume that the dynamic process of $H_t$ satisfies the second equation in (\ref{basic}), and we have observations $H_0,H_1,\cdots, H_T$. Let $h\triangleq(h_1,h_2,\cdots,h_T)'$ be the log-return process, then we have the likelihood function
\begin{align*}
p(h|\mu_h,\sigma_h^2)
= \frac{1}{(2\pi)^{\frac{T}{2}}(\sigma_h^2)^{\frac{T}{2}}}\prod\limits_{t=1}^T 
\exp \left\{ {-\frac{[h_t-(\mu_h-\frac{1}{2}\sigma_h^2)]^2}{2\sigma_h^2}}\right \}.
\end{align*}

%

We set noninformative Reference prior density \cite{Berger(1989)} for $\mu_h$ and $\sigma_h^2$
\begin{align}\label{yprior}
p(\mu_h,\sigma_h^2)\propto\frac{1}{\sigma_h}.
\end{align}
Given the return data $h_t,\,t=1,2,\cdots,T$ and the prior specified above, by Bayesian theorem, we have the joint posterior for $\mu_h$ and $\sigma_h^2$
\begin{align}\label{1prior}
p(\mu_h,\sigma_h^2|h)
&\propto p(h|\mu_h,\sigma_h^2)\cdot p(\mu_h,\sigma_h^2)
\nonumber \\& \propto \frac{1}{(\sigma_h^2)^{\frac{T+1}{2}}}\prod\limits_{t=1}^T 
 \exp \left\{ {-\frac{[h_t-(\mu_h-\frac{1}{2}\sigma_h^2)]^2}{2\sigma_h^2}}\right \}
\nonumber \\& \propto \frac{1}{(\sigma_h^2)^{\frac{T+1}{2}}}  \exp \Bigg \{ {-\frac{T[\mu_h-\bar{h}-\frac{1}{2}\sigma_h^2]^2+\sum\limits_{t=1}^T {h_t}^2-T\bar{h}^2}{2\sigma_h^2}}\Bigg\},
\end{align}
where $\bar{h}$ is the sample mean of $h_t$. 

To further infer parameters entering the dynamics of $X_t|H_t$, it is natural to take the posterior from $H_t$ derived above in (\ref{1prior}) as a source of prior information.
Note that
$$x_t|h_t\thicksim N\Bigg[(\mu_x-\displaystyle\frac{1}{2}\sigma_x^2)+\rho\displaystyle\frac{\sigma_x}{\sigma_h}\,(h_t
-(\mu_h-\displaystyle\frac{1}{2}\sigma_h^2)),\,(1-\rho^2)\sigma_x^2\Bigg],$$
with $x_t$ and $h_t$ defined in equation (\ref{return}). Let $x\triangleq(x_1,x_2,\cdots,x_T)'$ 
be the returns vector, we obtain the likelihood function for observations $x$ and $h$ in terms of parameters $\mu_x,\mu_y, \sigma_x,\sigma_h$ and $\rho$, which is expressed by
\begin{align}\label{like}
\nonumber p(x|h, \mu_x,\mu_h,&\sigma_x,\sigma_h,\rho)= \frac{1}{(2\pi)^{\frac{T}{2}}(1-\rho^2)^{\frac{T}{2}}(\sigma_x)^T}
\\&\times\prod\limits_{t=1}^T \exp\Bigg\{-\frac{[x_t-(\mu_x-\frac{1}{2}\sigma_x^2)-\rho\frac{\sigma_x}{\sigma_h}(h_t-(\mu_h-\frac{1}{2}\sigma_h^2))]^2}{2\sigma_x^2(1-\rho^2)}\Bigg\}.
\end{align}
Here, the joint prior of $(\mu_x,\mu_h,\sigma_x,\sigma_h,\rho)$ is given by
\begin{align}\label{jointprior}
p(\mu_x,\mu_h,\sigma_x,\sigma_h,\rho)\propto\! \frac{1}{(1-\rho^2)\sigma_x\sigma_h^{T}}
\exp\Bigg \{ {-\frac{T[\mu_h-\bar{h}-\!\frac{1}{2}\sigma_h^2]^2+\!\sum\limits_{t=1}^T {h_t}^2-\!T\bar{h}^2}{2\sigma_h^2}} \Bigg\}.
\end{align}
In particular, $p(\mu_x,\sigma_x,\rho|\mu_h,\sigma_h)\propto\dfrac{1}{(1-\rho^2)\sigma_x}$ obtained by using the Reference prior rule \cite{Berger(1989)}.
Then, combining prior density function (\ref{jointprior}) with the corresponding likelihood function (\ref{like}), we have the joint posterior
\begin{align}\label{jp}
\nonumber p \Big(\mu_x,\mu_h,&\sigma_x,\sigma_h,\rho \Big|x,h \Big)\propto p\left(x|h,\mu_x,\mu_h,\sigma_x,\sigma_h,\rho\right)p(\mu_x,\mu_h,\sigma_x,\sigma_h,\rho)
\nonumber \\& \propto \frac{1}{(1-\rho^2)^{\frac{T}{2}+1}\sigma_x^{T+1}\sigma_h^{T}}
\exp \Bigg \{ {-\frac{T[\mu_h-\bar{h}-\frac{1}{2}\sigma_h^2]^2+\sum\limits_{t=1}^T (h_t-\bar{h})^2}{2\sigma_h^2}} \Bigg\}
\nonumber \\& \quad \times\prod\limits_{t=1}^T \exp\Bigg\{-\frac{\left[x_t\!-\!(\mu_x\!-\!\frac{1}{2}\sigma_x^2)\!-\!\rho\frac{\sigma_x}{\sigma_h}(h_t\!-\!(\mu_h\!-\!\frac{1}{2}\sigma_h^2))\right]^2}{2\sigma_x^2(1\!-\!\rho^2)}\Bigg\}.
\end{align}

In option pricing applications, the uncertainty is independent of parameter $\mu_x,\mu_h$ under the risk-neutral measure, so we are only interested in the posteriors of $\sigma_x,\sigma_h, \rho$ here. 
Integrating (\ref{jp}) with respect to $\mu_x$ and $\mu_h$ yields the joint posterior density 
\begin{align}\label{fullpost}
\nonumber p(&\sigma_x,\sigma_h,\rho\Big|x,h) =\int \int p \Big(\mu_x,\mu_h,\sigma_x,\sigma_h,\rho \Big|x,h \Big) \,\mathrm{d}\mu_x\mathrm{d}\mu_h
\nonumber \\& \propto  \frac{1}{(1-\rho^2)^{\frac{T}{2}}\sigma_x^{T}\sigma_h^{T-1}}
 \exp \Bigg[-\frac{\sum\limits_{t=1}^{T}(x_t\!-\!\bar{x})^2}{2\sigma_x^2(1\!-\!\rho^2)}
-\frac{\sum\limits_{t=1}^{T}(h_t\!-\!\bar{h})^2}{2\sigma_h^2(1\!-\!\rho^2)}
-\frac{\rho(T\bar{x}\bar{h}-\sum\limits_{t=1}^{T}x_th_t)}{\sigma_x\sigma_h(1\!-\!\rho^2)}\Bigg]
\nonumber \\& \quad \times \Bigg\{  \int \int \frac{1}{\frac{2 \pi\,\sigma_x\sigma_h \sqrt{1\!-\!\rho^2}}{T}} 
 \exp\Bigg[-\frac{(\mu_x-\bar{x}-\frac{1}{2}\sigma_x^2)^2}{\frac{2\sigma_x^2(1-\rho^2)}{T}}-\frac{(\mu_h-\bar{h}-\frac{1}{2}\sigma_h^2)^2}{\frac{2\sigma_h^2(1-\rho^2)}{T}}
\nonumber \\&
\qquad \qquad +\rho \frac{(\mu_x-\bar{x}-\frac{1}{2}\sigma_x^2)(\mu_h-\bar{h}-\frac{1}{2}\sigma_h^2)}{\frac{\sigma_x\sigma_h(1-\rho^2)}{T}}\Bigg ] \mathrm{d}\mu_x\mathrm{d}\mu_h \Bigg \},
\end{align}
where $\bar{x}$ is the sample mean of $x_t$. The two-dimensional integration in the large brace is equal to 1. That is, 
\begin{align}
\nonumber p(\sigma_x,\sigma_h,\rho\Big|x,h)  \propto  \frac{1}{(1-\rho^2)^{\frac{T}{2}}\sigma_x^{T}\sigma_h^{T-1}}
 \exp \Bigg[-\frac{\sum\limits_{t=1}^{T}(x_t\!-\!\bar{x})^2}{2\sigma_x^2(1\!-\!\rho^2)}
-\frac{\sum\limits_{t=1}^{T}(h_t\!-\!\bar{h})^2}{2\sigma_h^2(1\!-\!\rho^2)}
-\frac{\rho(T\bar{x}\bar{h}-\sum\limits_{t=1}^{T}x_th_t)}{\sigma_x\sigma_h(1\!-\!\rho^2)}\Bigg],
\end{align}
 and we can verify that this posterior density of $\theta \triangleq(\sigma_x,\sigma_h,\rho)$ is a proper one. Then we obtain the conditional posterior densities
\begin{align}\label{c1}
p\left(\sigma_x \big|\sigma_h,\rho, x,h\right)\propto
\frac{1}{(\sigma_x)^{T}}
\exp\Bigg[-\frac{\sum\limits_{t=1}^{T}(x_t\!-\!\bar{x})^2}{2\sigma_x^2(1\!-\!\rho^2)}-\frac{\rho(T\bar{x}\bar{h}-\sum\limits_{t=1}^{T}x_th_t)}{\sigma_x\sigma_h(1\!-\!\rho^2)}\Bigg],
\end{align}
\begin{align}\label{c2}
p\left(\sigma_h \big|\sigma_x,\rho, x,h\right)\propto
\frac{1}{(\sigma_h)^{T-1}}
\exp\Bigg[-\frac{\sum\limits_{t=1}^{T}(h_t\!-\!\bar{h})^2}{2\sigma_h^2(1\!-\!\rho^2)}-\frac{\rho(T\bar{x}\bar{h}-\sum\limits_{t=1}^{T}x_th_t)}{\sigma_x\sigma_h(1\!-\!\rho^2)}\Bigg],
\end{align}
\begin{align}\label{c3}
p\left(\rho\big|\sigma_x,\sigma_h, x,h\right)\propto &
\frac{1}{(1\!-\!\rho^2)^{\frac{T}{2}}}
\exp\Bigg[-\frac{\sum\limits_{t=1}^{T}(x_t\!-\!\bar{x})^2}{2\sigma_x^2(1\!-\!\rho^2)}
\!-\!\frac{\rho^2\sum\limits_{t=1}^{T}(h_t\!-\!\bar{h})^2}{2\sigma_h^2(1\!-\!\rho^2)}
\!-\!\frac{\rho(T\bar{x}\bar{h}\!-\!\sum\limits_{t=1}^{T}x_th_t)}{\sigma_x\sigma_h(1\!-\!\rho^2)}\Bigg].
\end{align}
\par
The conditional posterior distributions (\ref{c1}), (\ref{c2}) and (\ref{c3}) are not kernels of classical distributions that we are familiar with. Thus, random samples can not be easily generated and the posterior simulation is imperative. MCMC are a class of algorithms that can be used to simulate from $p\left(\sigma_x,\sigma_h,\rho| x,h\right)$. In particular, the algorithm of Metropolis with-in Gibbs \cite{Besag(1995)} will be introduced in section 5 to construct Markov chains with stationary distributions $p\left(\sigma_x \big|\sigma_h,\rho, x,h\right)$, $p\left(\sigma_h \big|\sigma_x,\rho, x,h\right)$ and $p\left(\rho\big|\sigma_x,\sigma_h, x,h\right)$, respectively.

\section{\large{Pricing Quanto options using Bayesian predictive densities}}
In the Quanto option contract, the holder of the option is exposed to risks from the exchange rate $H_t$ as well as the foreign underlying asset $X_t$, and the correlation $\rho$ between $X_t$ and $H_t$. Since Quanto options prices are functions of parameters $\sigma_x, \sigma_h, \rho$, we can compute the prices using Monte Carlo methods on the basis of the posterior results of parameters obtained in the last section.

Here, we draw inspiration from Rombouts et al.\,\cite{Rombouts(2014)} and Bauwens et al.\,\cite{BauwensandLubrano(2002)} and explain how to price four types of Quanto options with payoff functions $F^{(1)}$, $F^{(2)}$, $F^{(3)}$ and $F^{(4)}$ defined previously in a Bayesian prediction framework with the uncertainty of parameters $\sigma_x, \sigma_h, \rho$ being considered. Unlike these previous studies, we predict the asset price $X$ and the exchange rate $H$ at the same time, and update the parameters $\sigma_x, \sigma_h, \rho$ step by step. Suppose that we have observations $H_0,H_1,\cdots,H_T$ and $X_0,X_1,\cdots,X_T$, we aim to predict the time-$(T+j), \,j\!=\!1,2,\cdots s$ prices of Quanto options with maturity $T+s$ using the available observations of underlying asset prices.

In the case of $\circled{1}$, from the risk-neutral valuation principle, the time-$(T\!+\!j)$ theoretical price of Quanto option with payoff $F^{(1)}$ is given by
\begin{align}\label{v1}
\nonumber V^{(1)}(&X,H,s\!-\!j)=e^{-r_d(s-j)}\,\mathbb{E}_{\mathbb{Q}_d}\left[\max(H_{T+s}X_{T+s}-K_d,0)\Big|\mathcal{F}_{T}\right]
\\&=e^{-r_d(s-j)}\,\int_0^\infty \max(H_{T+s}X_{T+s}-K_d,0)p(X_{T+s},H_{T+s})\,\mathrm{d}X_{T+s}\,\mathrm{d}H_{T+s},
\end{align}
where $p(X_{T+s},H_{T+s})$ is the density of the asset price at maturity under the domestic risk neutral measure denoted by $\mathbb{Q}_d$.

An exact analytical expression of the integration in (\ref{v1}) is generally not possible, and the Monte Carlo simulation method will be numerically used to approximate the integration. 
The draws of $X_{T+j}$ and $H_{T+j}$ can be simulated by using the similar method in \cite{Rombouts(2014)}. More precisely, the returns of $X_{T+j}$ and $H_{T+j}$ are sampled from their respective predictive density, and then these returns are transformed into draws of $X_{T+j}$ and $H_{T+j}$ according to the relationship between asset prices and returns. 

Let $x_{T+j}\triangleq \ln \frac{X_{T+j}}{X_{T+j-1}}$, $h_{T+j}\triangleq\ln \frac{H_{T+j}}{H_{T+j-1}}$ and $\theta\triangleq(\sigma_x,\sigma_h,\rho)$. When $j\!=\!1$, the predictive density of $x_{T\!+\!1}$ and $h_{T\!+\!1}$ \cite{Rombouts(2014), BauwensandLubrano(2002)} under measure $\mathbb{Q}_d$ is given by
\begin{align}\label{predict}
p\left(x_{T+1},h_{T+1}\big|x,h\right)=\int p\left(x_{T+1},h_{T+1}\big|\theta,x,h\right) p\left(\theta\big|x,h\right)\,\mathrm{d}\theta,
\end{align}
where the posterior density $p\left(\theta|x,h\right)$ is derived in equation (\ref{fullpost}), and $p\left(x_{T\!+\!1},h_{T\!+\!1}|\theta,x,h\right)$ under measure $\mathbb{Q}_d$ is given by
\begin{align}\label{pdfd}
\nonumber p&(x_{T+1},h_{T+1}\big|\theta,x,h) = \displaystyle\frac{1}{2 \pi\,\sigma_x\sigma_h \sqrt{1\!-\!\rho^2}}
 \exp \Bigg\{-\displaystyle\frac{\left[x_{T\!+\!1}-\left(r_f\!-\!\rho\sigma_x\sigma_h\!-\!\frac{1}{2}{\sigma_x}^2\right)\right]^2}{2 {\sigma_x}^2(1\!-\!\rho^2)}
\\ &-\displaystyle\frac{\left[h_{T\!+\!1}\!-\!\left(r_d\!-\!r_f\!-\!\frac{1}{2}{\sigma_h}^2\right)\right]^2}{2 {\sigma_h}^2(1\!-\!\rho^2)}
+\displaystyle \frac{\left[x_{T\!+\!1}\!-\!(r_f\!-\!\rho\sigma_x\sigma_h\!-\!\frac{1}{2}{\sigma_x}^2)\right]\left[h_{T\!+\!1}\!-\!(r_d\!-\!r_f\!-\!\frac{1}{2}{\sigma_h}^2)\right]\rho}{\sigma_x \sigma_h (1\!-\!\rho^2)}\Bigg\}.
\end{align}
Then we approximate (\ref{predict}) by
\begin{align*}
{p}\left(x_{T+1},h_{T+1}\big|x,h\right)\simeq\frac{1}{K\!-\!K_0}\sum_{k=K_0+1}^K  p\left(x_{T+1},h_{T+1}\big|\theta^{(k)},x,h\right),
\end{align*}
where $\theta^{(k)}$ is derived from the posterior (\ref{fullpost}), and the first $K_0$ draws are dropped to eliminate the initial value effect.
\par
When $j\!\geq\!2$, conditioning on the information of drawing $x_{T+1}, h_{T+1}$, we use similar procedures \cite{Rombouts(2014), BauwensandLubrano(2002)}  to generate $x_{T+2},h_{T+2},\cdots,x_{T+s},h_{T+s}$ step by step. Therefore, the predictive density of $x_{T+j}, h_{T+j}$ under measure $\mathbb{Q}_d$, is given by
\begin{align}\label{predd}
\nonumber p(x_{T+j},&h_{T+j}\big| x,h)=\int\int \int \cdots \int\int p\left(x_{T+j},h_{T+j}\big|\theta^{(k)}, x_{T+j-1},h_{T+j-1},\cdots, x,h\right)
\nonumber \\& \times p\left(x_{T+j-\!1},h_{T+j-\!1}\big|\theta^{(k)}, x_{T+j-2},h_{T+j-2},\cdots,x_{T+1}, h_{T+1},x,h\right) \cdots
\nonumber \\ & \times p\left(x_{T+1},h_{T+1}\big|\theta^{(k)}, x,h\right)p\left(\theta^{(k)}\big|x,h\right)\,\mathrm{d}x_{T+1}\,\mathrm{d}h_{T+1}\cdots\,\mathrm{d}x_{T+j-1}\,\mathrm{d}h_{T+j-1}\mathrm{d}\theta^{(k)}.
\end{align}
The high-dimensional integration is a great challenge, but we can approximate it in numerical simulations, that is, $x_{T+j}$, $h_{T+j}$ are simulated from their joint density with the parameters being sampled from their posterior densities (\ref{fullpost}). 
Once the return sequences $x_{T+1},x_{T+2}\cdots,x_{T+s}$ and $h_{T+1},h_{T+2}\cdots, h_{T+s}$ are simulated, we respectively approximate $X_{T+s}$ and $H_{T+s}$ by Monte Carlo method
$${X}_{T+s}\simeq\frac{1}{K\!-\!K_0} \sum_{k=K_0+1}^K X_T\,\exp\left(\sum_{j=T+1}^{T+s} x_{j}^{(k)}\right), $$
$${H}_{T+s}\simeq\frac{1}{K\!-\!K_0} \sum_{k=K_0+1}^K H_T\,\exp\left(\sum_{j=T+1}^{T+s} h_{j}^{(k)}\right).$$
Therefore, the price of Quanto options with payoff function $F^{(1)}$ can be further approximated by
\begin{align*}
{V}^{(1)}(X,H,s\!-\!j)\simeq & \,\,e^{-r_d(s-j)}\,\frac{1}{K\!-\!K_0}
\\ & \times \sum_{k=K_0+1}^K \left[\max\left(X_{T} \exp\left(\sum_{j=T\!+\!1}^{T\!+\!s} x_j^{(k)}\right)H_T\,\exp\left(\sum_{j=T+1}^{T+s} h_{j}^{(k)}\right)\!-\!K_f,0\right)\right],
\end{align*}
where $x_{T+j}^{(k)}$ and $h_{T+j}^{(k)}$ are simulated from $p\left(x_{T+j},h_{T+j}\big|\theta^{(k)},x,h\right)$, and $\theta^{(k)}$ is simulated from the corresponding posterior density based on a sequence of adjusted returns samples $x_1,x_2,$ $\cdots x_{T+j-1}$ and $h_1,h_2,\cdots h_{T+j-1}$.

By using the similar procedure of approximating $V^{(1)}(X,H,s-\!j)$, we approximate the prices of quanto options with other three forms of payoff functions as following.

In the case of $\circled{2}$, the foreign asset is settled in foreign currency, and is finally converted to domestic currency using prevailing exchange rate. Here, we consider the distribution of $x_{T+j}$ and $h_{T+j}$ under the domestic risk neutral measure $\mathbb{Q}_d$, 
then
\begin{align}\label{v2}
\nonumber V^{(2)}(&X,H,s\!-\!j)=e^{-r_d(s-j)}\, \mathbb{E}_{\mathbb{Q}_d} \left[H_{T+s}\max(X_{T+s}-K_f,0)\Big|\mathcal{F}_{T}\right]
\nonumber\\&=e^{-r_d(s-j)}\,\int_0^\infty H_{T+s} \max(X_{T+s}-K_d,0)p(X_{T+s},H_{T+s})\,\mathrm{d}X_{T+s}\mathrm{d}H_{T+s}
\nonumber\\&\simeq e^{-r_d(s-i)}\,\frac{1}{K\!-\!K_0}
\nonumber\\&\quad \times \sum_{k\!=K_0\!+1}^K \left\{H_T\,\exp\left(\sum_{i=T\!+1}^{T\!+j} h_{i}^{(k)}\right)\left[\max\left(X_{T} \exp\left(\sum_{j=T\!+\!1}^{T\!+\!s} x_j^{(k)}\right)\!-\!K_d,0\right)\right]\right\}.
\end{align}
Here, $x_{T+j}^{(k)}$ and $h_{T+j}^{(k)}$ are simulated from their joint probability density under the measure $\mathbb{Q}_d$ with the parameters being sampled from their posterior densities (\ref{fullpost}).
Although parameters $\rho$ and $\sigma_h$ do not enter into the corresponding pricing formula, we consider their randomness when predict exchange rate prices at time $T+s$.

In the case of $\circled{3}$, we have
\begin{align}\label{v4}
\nonumber V^{(3)}(&X,H,s\!-\!j)=e^{-r_d(s-j)}\,\mathbb{E}_{\mathbb{Q}_d}\left[H_{fix}\max(X_{T+s}-K_f,0)\Big|\mathcal{F}_{T}\right]
\nonumber \\&=e^{-r_d(s-j)}\,\int_0^\infty H_{fix}\max(X_{T+s}-K_f,0)p(X_{T+s})\,\mathrm{d}X_{T+s}
\nonumber \\&\simeq e^{-r_d(s-i)}\,\frac{1}{K\!-\!K_0}\sum_{k=K_0+1}^K \left[H_{fix}\max\left(X_{T} \exp\left(\sum_{j=T\!+\!1}^{T\!+\!s} x_j^{(k)}\right)-K_f,0\right)\right].
\end{align}
Here, $x_{T+j}^{(k)}$ are simulated from its probability density $N(r_f-\rho\sigma_h\sigma_x-\frac{1}{2}\sigma_x^2,\sigma_x^2)$ under the measure $\mathbb{Q}_d$, and the parameter $\theta^{(k)}=(\rho^{(k)},\sigma_x^{(k)},\sigma_h^{(k)})$ is simulated from the joint posterior density (\ref{fullpost}).

In the case of $\circled{4}$, the settled asset is exchange rate. 
We have
\begin{align}\label{v3}
\nonumber V^{(4)}(&X,H,s\!-\!j)=e^{-r_d(s-j)}\,\mathbb{E}_{\mathbb{Q}_d}\left[X_{T+s}\max(H_{T+s}-K_H,0)\Big|\mathcal{F}_{T}\right]
\\&=e^{-r_d(s-j)}\,\int_0^\infty X_{T+s}\max(H_{T+s}-K_H,0)p(X_{T+s},H_{T+s})\,\mathrm{d}X_{T+s}\,\mathrm{d}H_{T+s}
\nonumber\\&\simeq e^{-r_d(s-i)}\,\frac{1}{K\!-\!K_0}
\nonumber\\&\quad \times \sum_{k=K_0+1}^K \left\{X_T\,\exp\left(\sum_{j=T+1}^{T+s} x_{j}^{(k)}\right)\left[\max\left(H_{T} \exp\left(\sum_{j=T\!+\!1}^{T\!+\!s} h_j^{(k)}\right)-K_H,0\right)\right]\right\}.
\end{align}
Here, $h_{T+j}^{(k)}$ and $x_{T+j}^{(k)}$ are simulated from their joint probability density under the measure $\mathbb{Q}_d$ with the parameters sampling from their posterior densities (\ref{fullpost}).

Note that, for example in the expression (\ref{predd}), the parameters $\sigma_x,\sigma_h,\rho$ are updated in each time step, i.e.\,the predictive data are further coupled with sample data to make Bayesian inference on parameters. This approach gives a more reasonable explanation for the uncertainty of unknown parameters than those do not update parameters step by step. However, this approach also comes at the cost of computation burden in the numerical study.
\section{\large Numerical simulations}
In this section, we perform empirical studies
to assess the validity of the method established as above by comparing it with the MLE method and the Bayesian method with conjugate informative priors. Our simulations involve two stages, sampling from the posterior density of parameters and further predicting option prices using the posterior results of parameters as inputs. 

In the first stage, as expressed in Eqs.\,(\ref{c1}), (\ref{c2}) and (\ref{c3}), the three conditional posteriors used for parameters estimations do not take forms of densities that are convenient to draw from, leading to the fact that the Gibbs sampling algorithm is not directly available anymore. However, the Metropolis-Hastings algorithm \cite{Muller(1991b)} has been shown to be valid for the posterior conditionals used in the Gibbs sampling algorithm, i.e.\,the Metropolis with-in Gibbs algorithm \cite{Besag(1995)}. Algorithm 1 shows the detailed steps of the algorithm used in our computation.
\begin{algorithm}[H]
\caption{Metropolis with-in Gibbs algorithm} 
1. For $i=1, 2, 3$, choose candidate density $q_i(\theta_i)$ and initial value $\theta_i^{(0)}$ for each parameter.
\vskip 1mm
For $k=1, 2, \cdots, K:$\\
2. Take a candidate draw $\theta_i^*$ from $q_i(\theta_i)$, \\
3. Calculate the acceptance probability
$$\alpha(\theta_i^{(k-1)},\theta_i^*)=\min\left\{1,\frac{p\left(\theta_i^* \Big|\theta_{\thicksim i}^{(k-1)}, x,h\right)}{p\left(\theta_i^{(k-1)} \Big|\theta_{\thicksim i}^{(k-1)}, x,h\right)}\frac{q_i(\theta_i^{(k-1)})}{q_i(\theta_i^*)}\right\}, $$
4. If $\alpha(\theta_i^{(k-1)},\theta_i^*)<u$, then set $\theta_i^{(k)}=\theta_i^{*}$, else $\theta_i^{(k)}=\theta_i^{(k-1)}$, where $u$ is a random draw from the uniform distribution $U(0,1).$
\end{algorithm}
\noindent where the subscript $i$ of $\theta_i$ denotes the $i$th element of vector $\theta=(\sigma_x, \sigma_h, \rho)$, $\theta_{\thicksim i}^{(k-1)}$ denotes draws generated in the $(k-1)$th step for elements in vector $\theta$ except the $i$th one, the superscripts $(k\!-\!1)$ and {*} denote the previous draw and current candidate draw for $\theta_i$, respectively. In this algorithm, the candidate density $q_i$ should be carefully chosen and a good choice for $q_i$ can generate better Markov chains that converge more quickly and efficiently to the desired distribution. We especially consider various types of candidate density functions for comparison, including the inverse Gamma (IG), the truncated Normal (TN), the truncated Student's $t$ (TT) and the Normal (N) densities.

The underlying data used for parameter estimations are daily closing levels of S\&P 500 index and the middle-rate of EUR-USD exchange rate, covering from Jan.\,6, 2011 to Oct.\,30, 2018. Fig.\,1 (a) and (b) plot the sample path of index returns and exchange rate returns.
\begin{figure}[H]
  \centering
  \subfigure[S\&P 500 index]{
  \includegraphics[height=4.45cm,width=7cm]{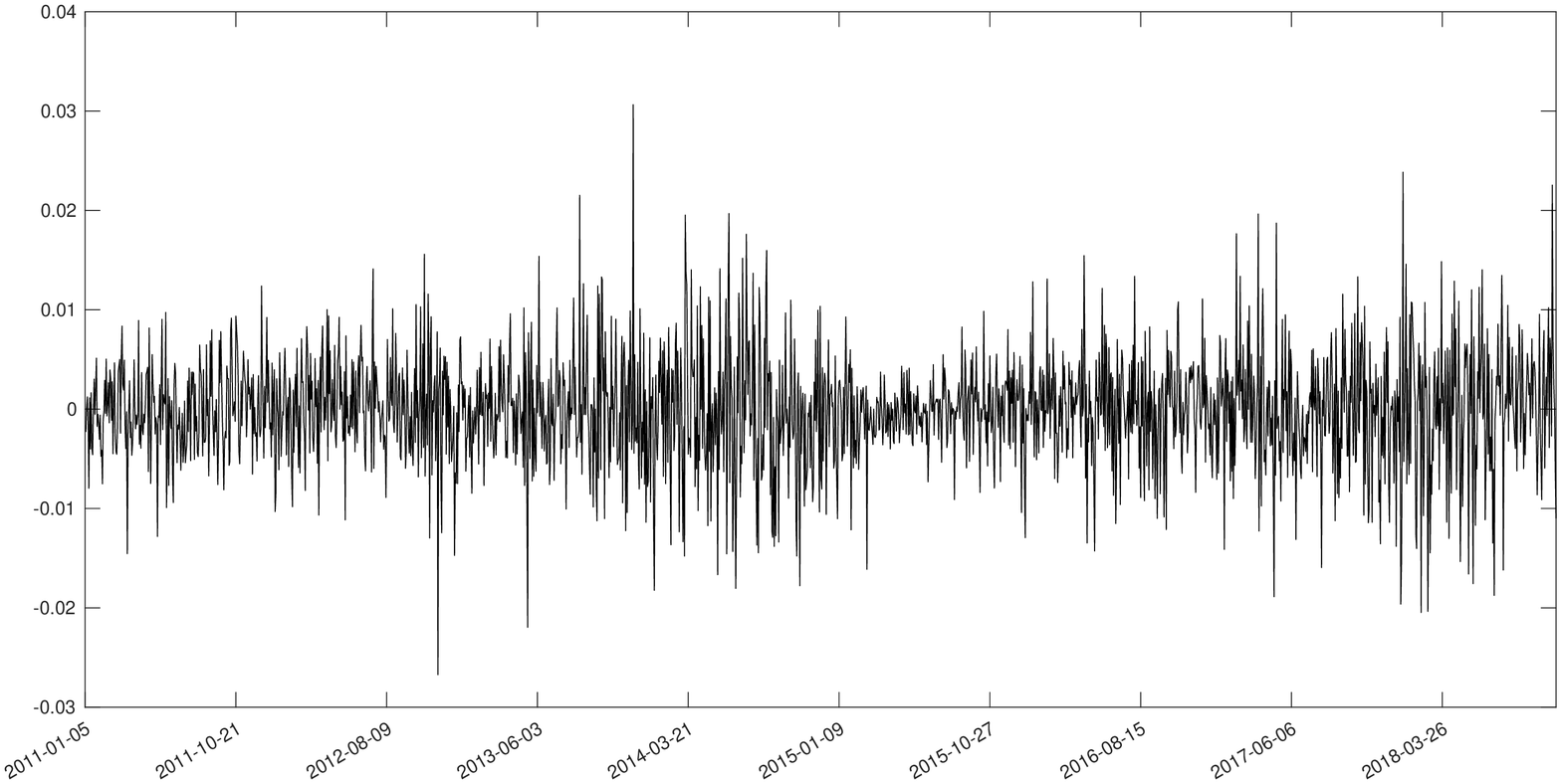}}
  \subfigure[EUR-USD exchange rate]{
  \includegraphics[height=4.45cm,width=7cm]{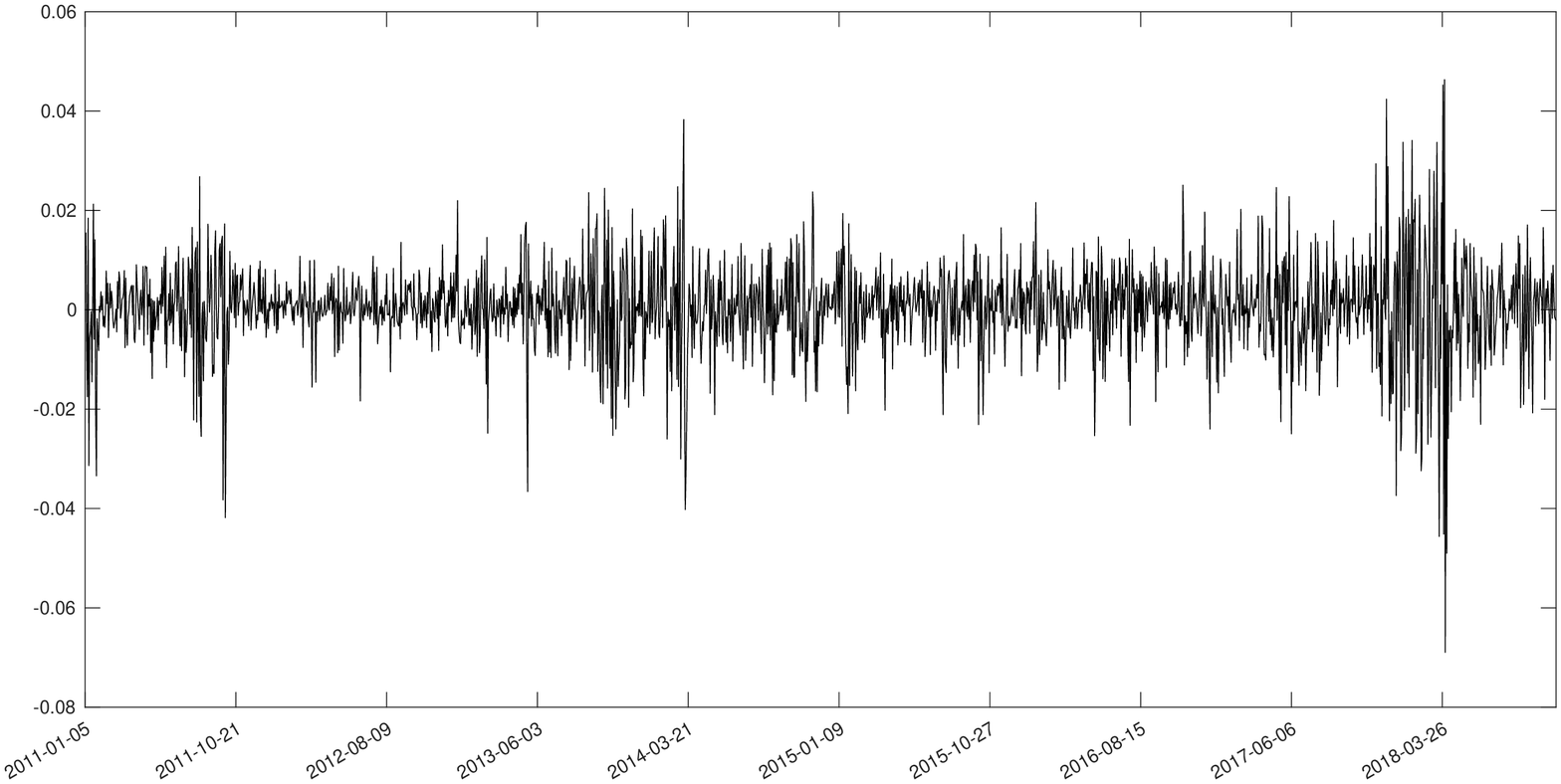}}
  \caption{Sample data for the log returns of S\&P 500 index and EUR-USD exchange rate}
\end{figure}

To remove the initial value effect, we take $K\!=\!300000$ and $K_0\!=\!100000$ in each simulation. We examine the posterior simulations across different sample sizes, but only present parameter results when the sample size is 140. Convergence of the Markov Chain generated by the Metropolis with-in Gibbs algorithm is diagnosed by two statistics `NSE' and `CD' \cite{Geweke(1992)}. 
\begin{center}
\small
\topcaption{Parameter estimations with various candidate densities}
\centering
\newcommand{\tabincell}[2]{\begin{tabular}{@{}#1@{}}#2\end{tabular}}
\renewcommand\arraystretch{1}
\begin{supertabular}{p{1.3cm}<{\centering}p{1.3cm}<{\centering}p{1.6cm}<{\centering}p{3cm}<{\centering}p{1.5cm}<{\centering}p{1.8cm}<{\centering}p{1.5cm}<{\centering}}\toprule   
  $ $              &Mean              &Std.dev.              & 95\% HPDI                    &NSE                                          &CD        \\ \midrule
  TTN             &                                                                                                                                          \\ 
 \midrule
$\sigma_x$    &0.0057           &8.1582$\times\!10^{-4}$  &[0.0051, 0.0062]     &1.7874$\times\!10^{-6}$             &0.15248         \\
$\sigma_h$    &0.0035           &0.0015                            &[0.0012, 0.0058]     &3.4259$\times\!10^{-6}$              &-0.69983         \\
$\rho$           &-0.0271           &0.0790                            &[-0.1334, 0.0782]     &1.7404$\times\!10^{-4}$              &0.49578         \\
\bottomrule
TNN             &                                                                                                                                            \\ 
\midrule
$\sigma_x$   &0.0062           &0.0047                        &[0.0005, 0.0152]        &1.0449$\times\!10^{-5}$            &-1.3207            \\
$\sigma_h$   &0.0025           &0.0019                        &[0.0002, 0.0062]       &4.2759$\times\!10^{-6}$            &0.62796             \\
$\rho$           &0.0392           &0.3476                        &[-0.5292, 0.6917]       &7.7716$\times\!10^{-4}$             &0.26118              \\
\bottomrule
IGN              &                                                                                                                                            \\ \midrule
$\sigma_x$   &0.0059           &9.9535$\times\!10^{-4}$  &[0.0048, 0.008]   &2.2257$\times\!10^{-6}$            &0.21235                  \\
$\sigma_h$   &0.0041           &7.8202$\times\!10^{-4}$  &[0.0029, 0.0055]  &1.7487$\times\!10^{-6}$            &-0.15433                  \\        
$\rho$           &0.0873           &0.5049                             &[-0.7406, 0.8697]  & 0.0011                                        &0.16212                      \\
\bottomrule
MNC                                                                                                                                                          \\ 
\midrule
$\sigma_x$       & 0.0072          & 3.046$\times\!10^{-4}$     &[0.0067, 0.0077]    & 6.8111$\times\!10^{-7}$         &-   
\\
$\sigma_h$       & 0.0046          &1.9523$\times\!10^{-4}$   &[0.0043, 0.0049]     &4.3654$\times\!10^{-7}$         &-
\\
$\rho$              & -0.0374         & 0.0598                              &[-0.1356, 0.0612]     &1.3364$\times\!10^{-4}$         &-
\\
\bottomrule
MLE               &                                                                                                                                          \\
\midrule
$\sigma_x$    &0.0083  &- &- &- &-
\\
$\sigma_h$    &0.0025 &- &- &- &-
\\  
$\rho$            &-0.0391 &- &- &- &-
\\
\bottomrule
\end{supertabular}
\end{center}

Table 1 reports the parameter posterior results using the previous 140 daily S\&P index and EUR-USD data. 
For comparison, the Bayesian conjugate posterior (MNC) result and the MLE result are presented in Table 1 as well. The MLE method only provides a point estimation, while Bayesian method offers the posterior density that can be used as inputs for further price options with the uncertainty of parameters taken into account. 
In table 1, TTN denotes that the candidate densities $q_1$, $q_2$, $q_3$ are TT, TT and N density, respectively. TNN denotes that the candidate densities $q_1$, $q_2$, $q_3$ are TN, TN and N density, respectively. IGN denotes that the candidate densities $q_1$, $q_2$, $q_3$ are IG, IG and N density, respectively. `Mean', `Std.dev.' and `Acp.' represent the posterior mean, posterior standard deviation and acceptance probability, respectively. The 'NSE' and 'CD' values both show that the Markov chains generated from our algorithms are convergent under different settings for candidate densities. 

Clearly, in Table 1, the Bayesian estimation results of model parameters with various candidate densities are somewhat different from the MNC and MLE results, especially for the correlation parameter $\rho$.

Fig.\,2 plots the corresponding posterior densities of each unknown parameters with different candidate densities being used in the posterior sampling simulations. 

\begin{figure}[H]
\centering
 \subfigure[Marginal posterior densities of $\sigma_x$, $\sigma_h$ and $\rho$ under the IGN candidate densities.]{
  \includegraphics[height=2.8cm,width=15cm]{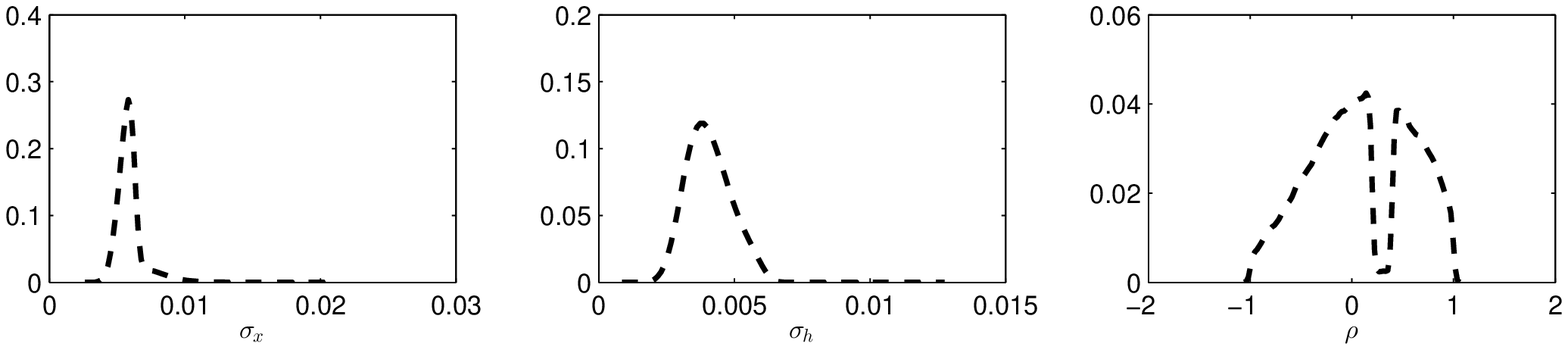}}
  \end{figure}
  \addtocounter{figure}{-1} 
  \begin{figure}[H]
\centering
  \subfigure[Marginal posterior densities of $\sigma_x$, $\sigma_h$ and $\rho$ under the TNN candidate densities.]{
  \includegraphics[height=2.8cm,width=15cm]{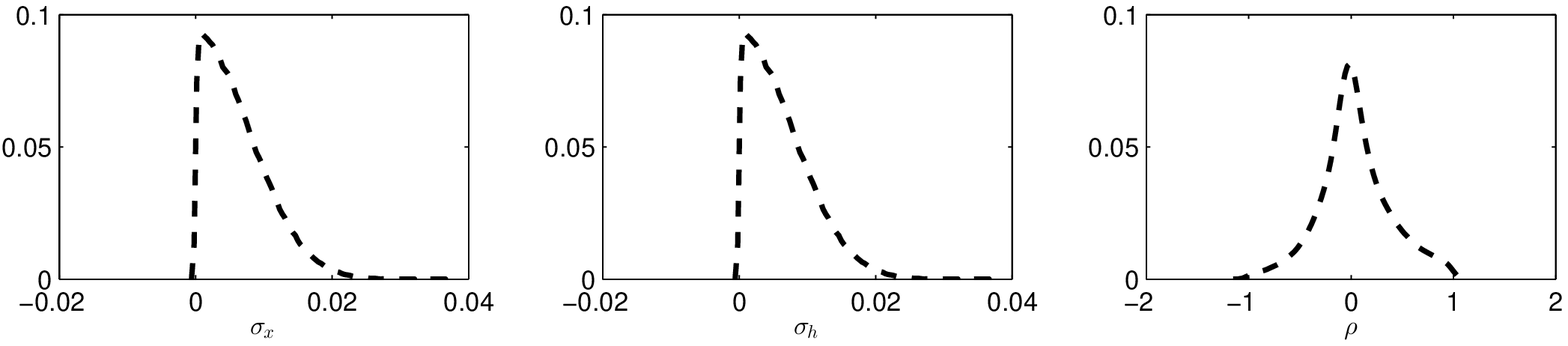}}
  \end{figure}
   \addtocounter{figure}{0} 
 \begin{figure}[H]
\centering
    \subfigure[Marginal posterior densities of $\sigma_x$, $\sigma_h$ and $\rho$ under the TTN candidate densities.]{
  \includegraphics[height=2.8cm,width=15cm]{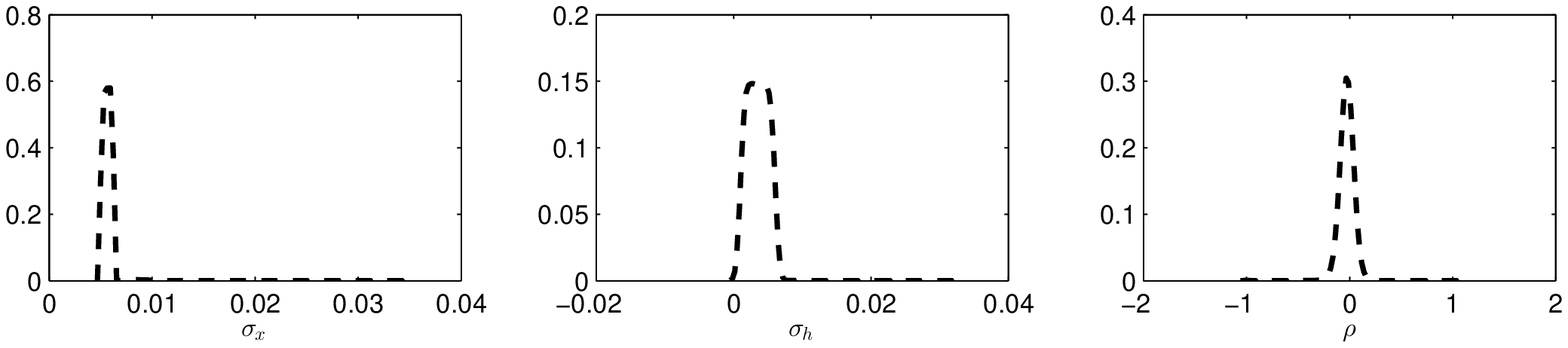}}
  \end{figure}
   \addtocounter{figure}{1} 
  \begin{figure}[H]
\centering
    \subfigure[Marginal posterior densities of $\sigma_x$, $\sigma_h$ and $\rho$ under the conjugate priors.]{
  \includegraphics[height=2.8cm,width=15cm]{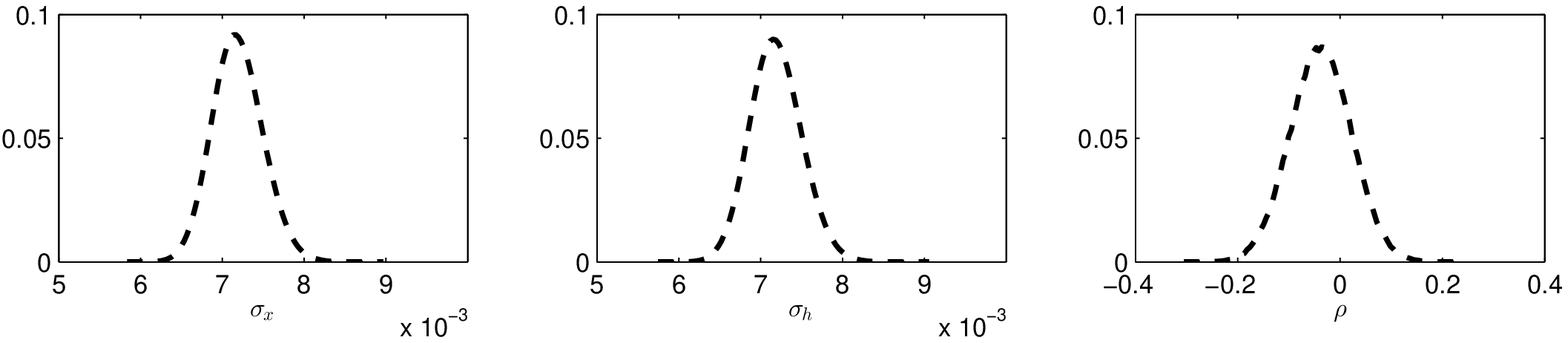}}
\caption{Comparison of the posterior densities of unknown parameters $\sigma_x$, $\sigma_h$ and $\rho$ under different candidate densities.}
\end{figure}

In the second stage, we apply our method to real market data obtained from the website ivolatility.\,com.
 Because of the absence of readily available data of Quanto option prices in real market, we follow the method in \cite{Fallahgoul(2018),Kim(2015)} and construct artificial Quanto option data implied by the real market data of S\&P 500 index option and EUR-USD exchange rate. For this reason, only the Quanto option with payoff function $F^{(3)}$ (fixed exchange rate foreign stock call struck in foreign currency) is discussed in the following examples examined in our numerical simulations. Let $QC(t,K_f,T)$ be the time-t S\&P 500 Quanto call option price and $C(t,K_f,T)$ the time-t S\&P 500 call option price, with strike $K_f$ and maturity $T$, then
\begin{equation}\label{construct}
QC(t,K_f,T)=e^{-r_d(T-t)}H_{fix}C(t,K_f,T),
\end{equation}
where the risk-free rates $r_d$ in domestic market and $r_f$ in foreign market are set according to the available LIBOR rate, $H_{fix}$ is set as 1 for simplicity. Therefore, the Quanto option is established in two steps \cite{Long(2015)}. At first, we take the call options on S\&P 500 index traded on Oct.\,31, 2018. Then, in the Euro zone, if one hopes to invest these options, the EUR-USD exchange rate is needed to convert the payoff settled in EUR. The option data has been filtered following the method in \cite{Steven(1982)} to exclude the options that might be exercised at some time before the maturity. 

We proceed to compare the simulated theoretical option prices with the real market data to examine whether our Bayesian method is efficient or not. In addition, 
we simulate theoretical BS price with implied volatility (BS-I) and BS price with historical volatility (BS-H) for comparison to indicate the advantage of Bayesian estimation method established in the previous sections of this paper. 
\begin{figure}[H]
\centering
  \includegraphics[height=7.5cm,width=15cm]{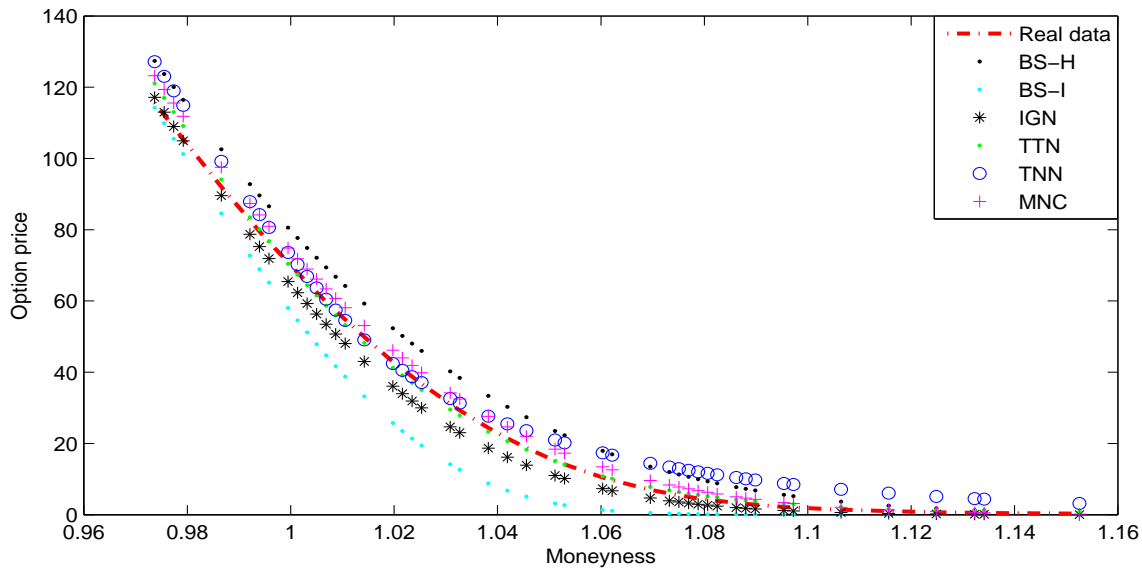}
  \caption{Comparison of real option data with simulated option data.}
\end{figure}
We particularly choose options with time-to-maturity  51 days and with different moneyness (defined as $\frac{K_f}{X_t}$) categories. Fig.\,3 plots the real market option data and the pricing results of options with varying moneyness categories under different methods for the purpose of comparison. As we can see from Fig.\,3, the Bayesian methods with TTN and IGN candidate densities have overall good agreements with real market data. The Bayesian method with TNN candidate density performs best among all the other methods for near the money options, but it is no longer valid for deep out-of the money options. In addition, we find that the Bayesian method constructed in this paper using non-informative priors has a better pricing performance than the pricing results using Bayesian conjugate priors, indicating that the adoption of conjugate priors may contain too much information as has been pointed by some researchers. Moreover, BS-I and BS-H models always underestimate and overestimate option prices in real market, respectively.
\begin{figure}[H]
\centering
\subfigure[IGN: mean=104.0741, NSE=0.12781, 99\% HPDI = \lbrack 91.5904, 120.4622\rbrack] 
{ 
  \includegraphics[height=3.8cm,width=6.2cm]{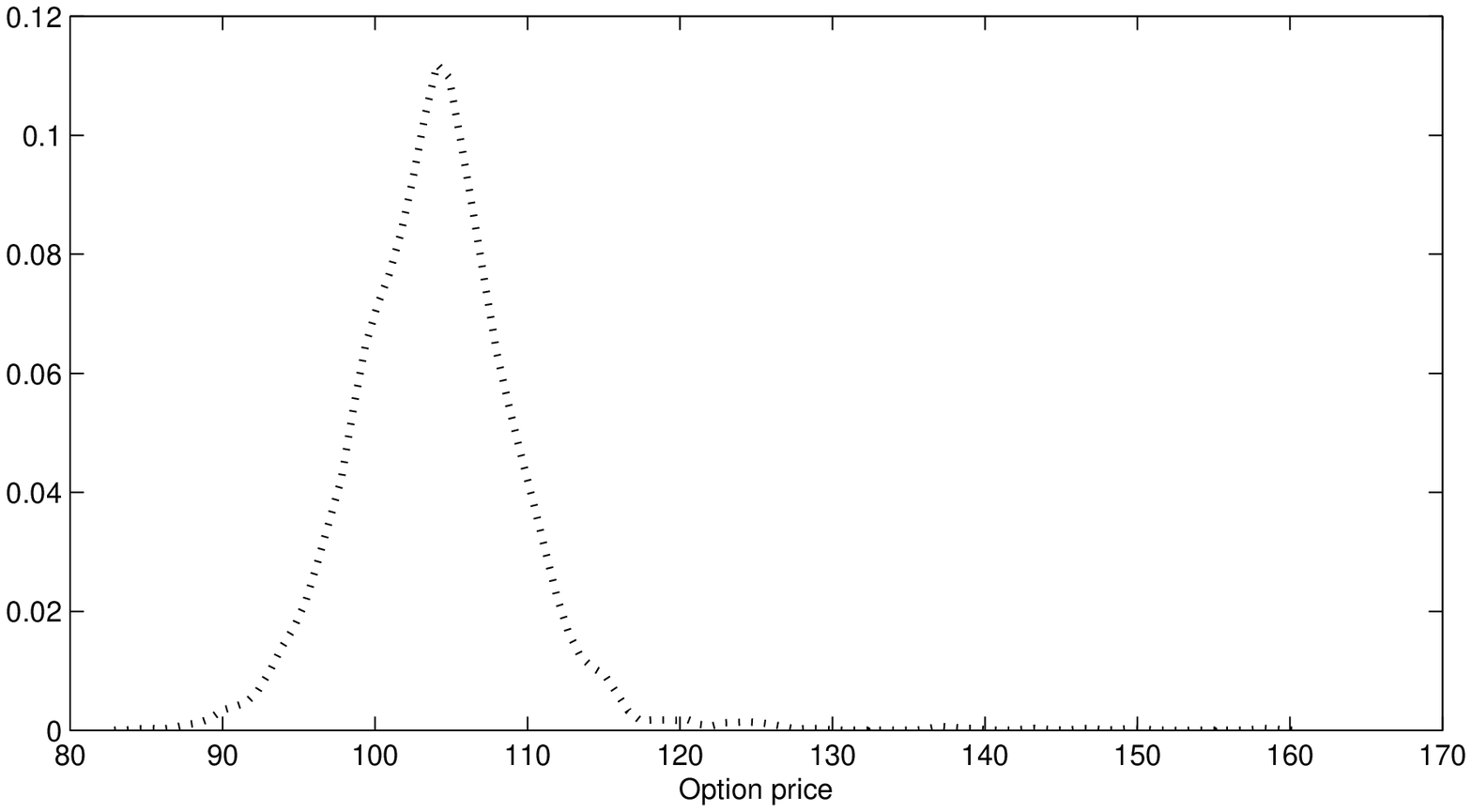}
}
\subfigure[TTN: mean=106.712, NSE=0.018313, 99\% HPDI = \lbrack 105.5674, 108.1829\rbrack]
{
  \includegraphics[height=3.8cm,width=6.2cm]{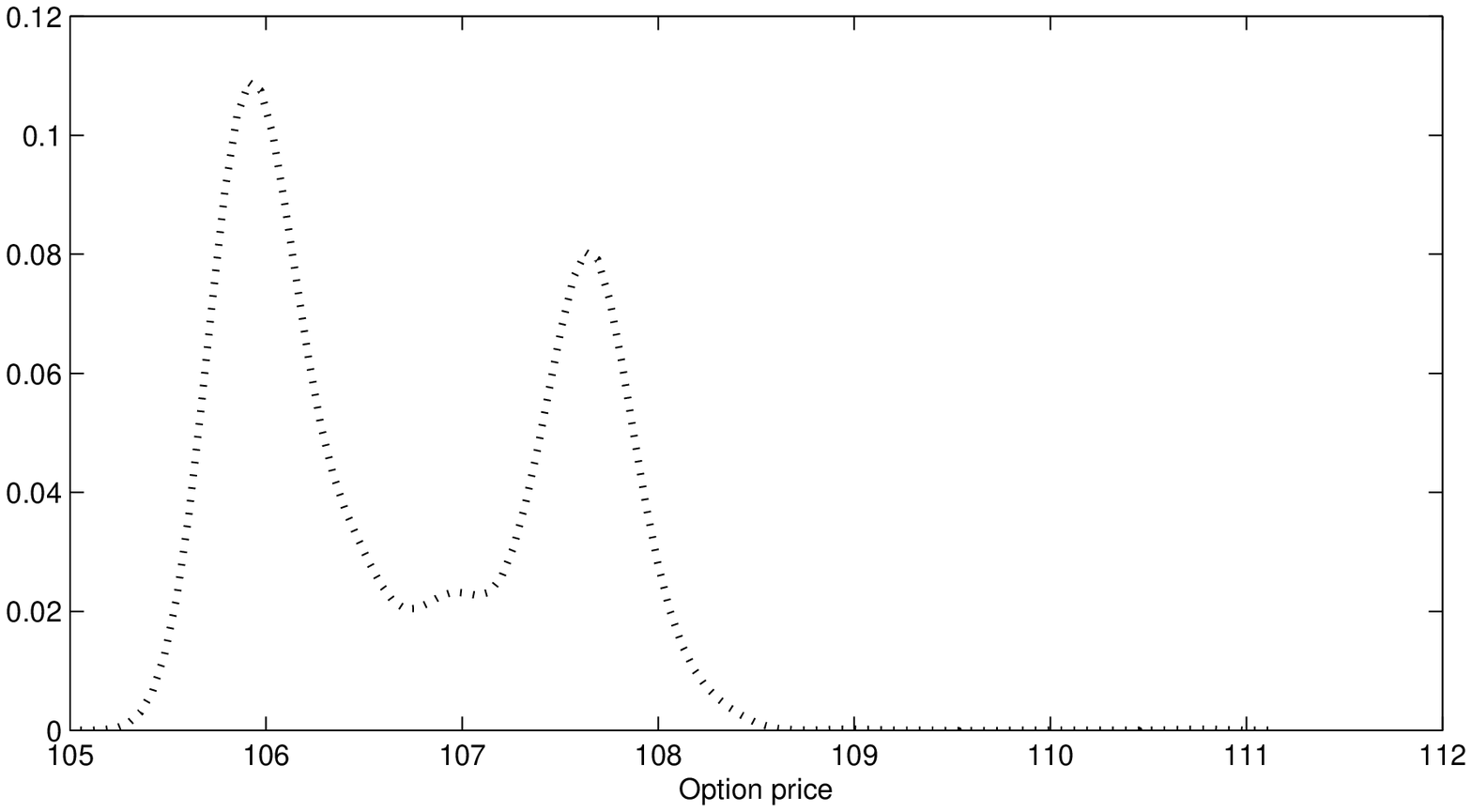}
  }
\subfigure[TNN: mean=108.1009, NSE=0.074932, 99\% HPDI = \lbrack 100.3838, 115.6493\rbrack]
{
  \includegraphics[height=3.8cm,width=6.2cm]{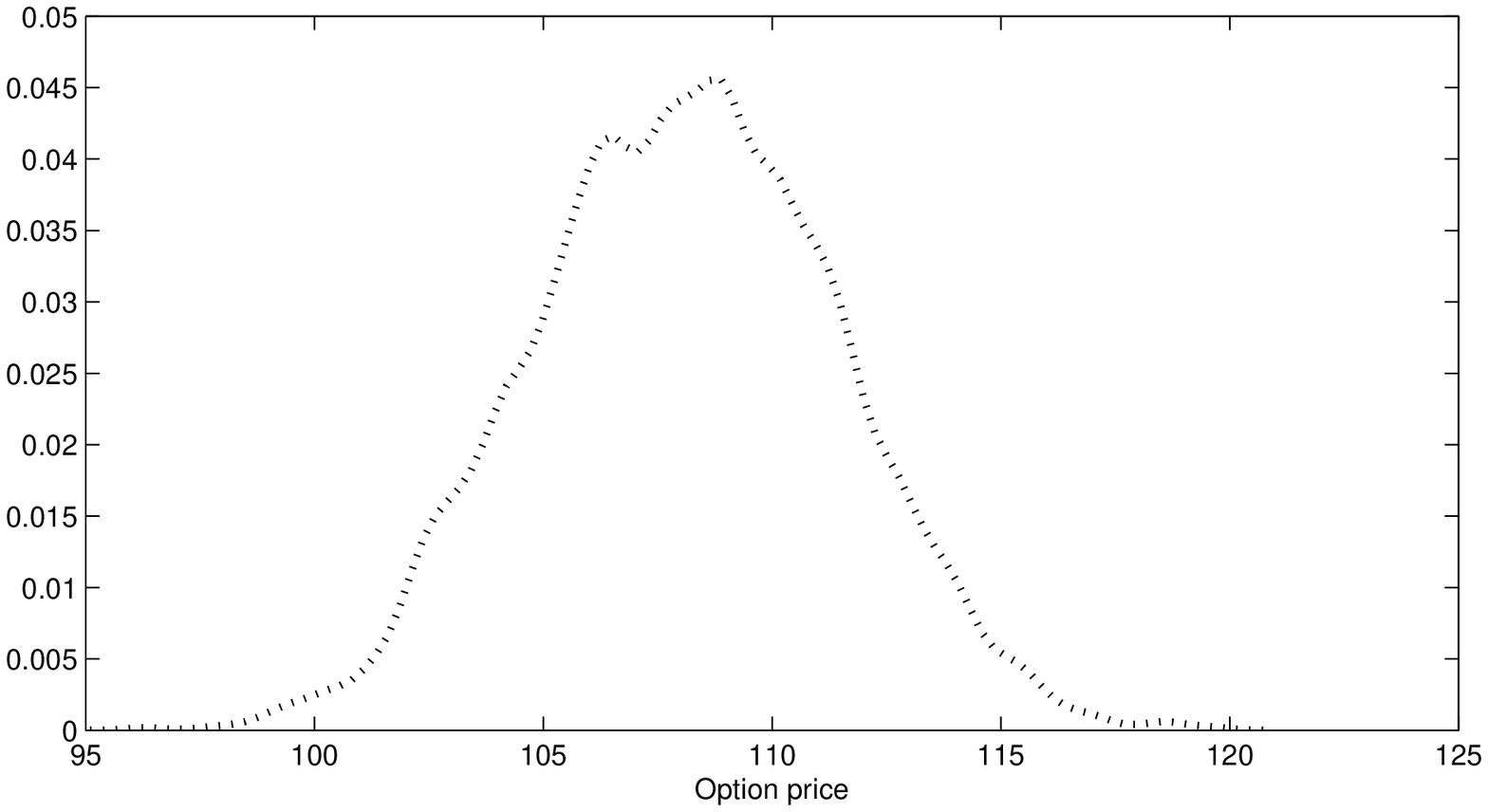}
  }
\subfigure[MNC: mean=114.297, NSE=0.001866, 99\% HPDI = \lbrack 114.095, 114.4969\rbrack] 
{
  \includegraphics[height=3.8cm,width=6.2cm]{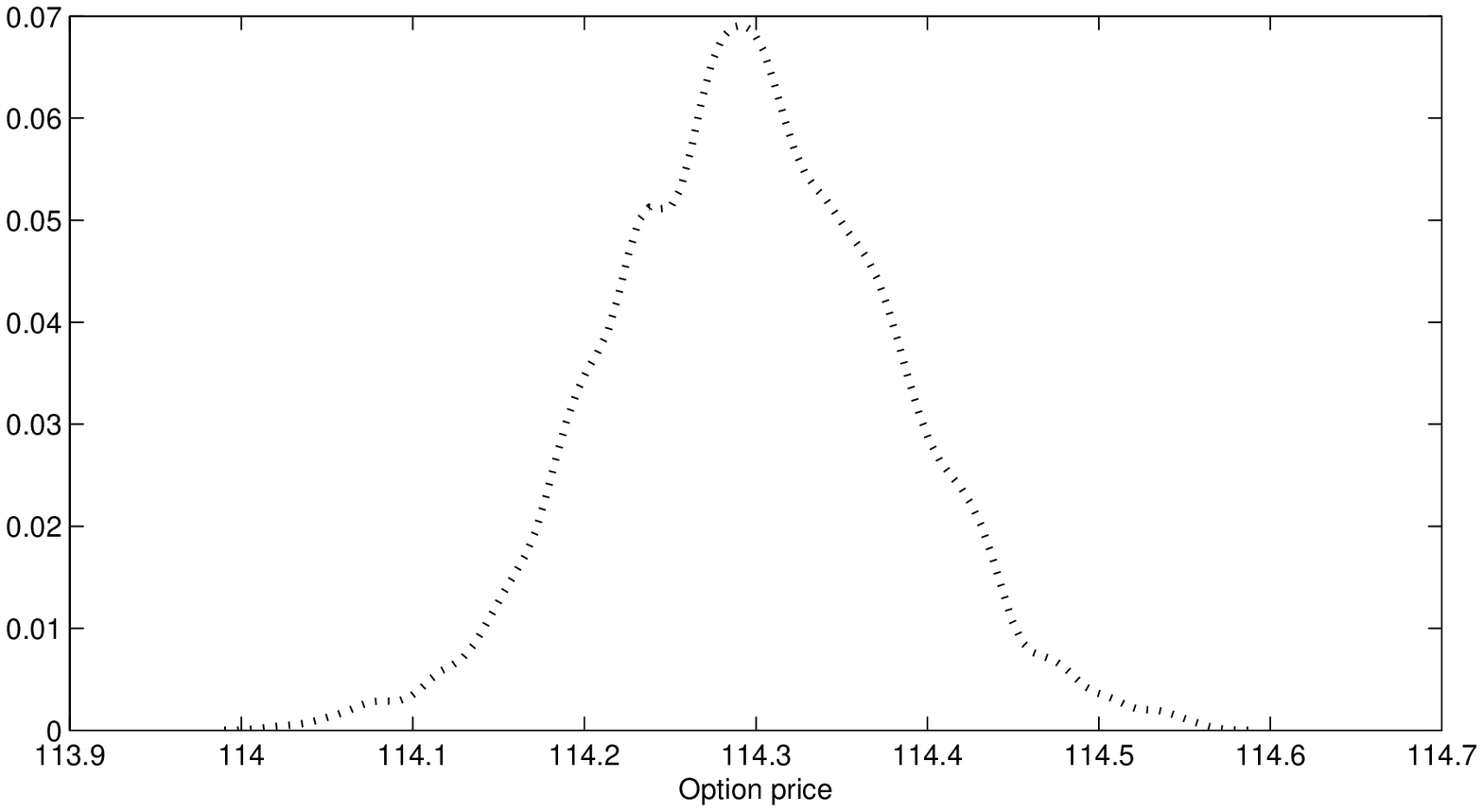}
  }
 \caption{Predictive densities of option prices across parameters posteriors with various candidate densities IGN(a), TTN(b), TNN(c), MNC(d), respectively.}
\end{figure}

We further demonstrate the effectiveness of predictive method constructed in section 4 in terms of pricing options, which updates parameters step by step. We consider the option with strike price 2655 and with time-to maturity 51 days, whose market price is 105.85. 
Fig.\,4 plots the predictive density of option price with posteriors of model parameters simulated from candidate densities IGN, TTN, TNN and MNC, respectively. The figures show how option prices are distributed, providing more useful information to traders than the classical estimation method that gives only a point forecast from the point view of a practitioner. 
We report the distribution features of predictive densities, including the mean, the NSE and the 99\% Highest Posterior Density Interval (HPDI) below each subfigure and find that IGN, TTN and TNN have better pricing performances than that of MNC, where the 99\% HPDI of IGN, TTN and TNN contain the true option value, showing again that the Bayesian method using non-informative priors might have more potentials than that of using informative conjugate priors.

The numerical simulation results presented above showed that the Bayesian inference method established in this paper has advantages in pricing options over the classical estimation method and the Bayesian method with conjugate priors, but this was only for one particular choice of exchange rate and for one fixed sample size. The last two aims of our numerical experiments is to check if the pricing performance remains valid with respect to (i) using different exchange rate processes for the construction of Quanto options; and (ii) the sample size of underlying assets data used for parameter estimations.

To demonstrate the effectiveness of the Bayesian method in this paper in terms of the pricing of constructed Quanto options with exchange rate processes between different currencies. We convert the European options on S\&P 500 index valued in US dollars to currencies settled in Euros, Pounds and Canadian dollars, respectively. Therefore, the exchange rate process between Euros and US dollars (EUR-USD), Pounds and US dollars (UK-USD), Canadian dollars and US dollars (CAN-USD) are required, and the Quanto options with exchange rates EUR-USD, UK-USD and CAN-USD are constructed following Eq.\,(\ref{construct}).
 \begin{center}
\small
\topcaption{Pricing performance for Quanto options constructed using exchange rates between different currencies under different moneyness.}
\centering
\newcommand{\tabincell}[2]{\begin{tabular}{@{}#1@{}}#2\end{tabular}}
\begin{supertabular}{p{2.2cm}p{1.5cm}<{\centering}p{2.6cm}<{\centering}p{2.6cm}<{\centering}p{2.6cm}<{\centering}}\toprule  
  Exchange Rate &Model          &ITM    &ATM             &OTM          \\\midrule
  $ $           &IGN      &\tabincell{c}{$0.01846$\\($0.215646$)}        &\tabincell{c}{$0.09083$\\($0.163425$)}      &\tabincell{c}{$0.39566$\\($0.05218$)}\\
  EUR-USD  &BS-I      &\tabincell{c}{$0.028699$\\($0.294447$)}       &\tabincell{c}{$0.22026$\\($0.132581$)}       &\tabincell{c}{$0.84966$\\($0.023036$)}\\
   $ $          &BS-H    &\tabincell{c}{$0.078091$\\($0.286977$)}       &\tabincell{c}{$0.152474$\\($0.219674$)}        &\tabincell{c}{$0.947846$\\($0.094999$)}   \\
\hline
  $ $          &IGN       &\tabincell{c}{$0.018556$\\($0.215611$)}       &\tabincell{c}{$0.09065$\\($0.163414$)}         &\tabincell{c}{$0.39583$\\($0.052171$)} \\
  UK-USD    &BS-I      &\tabincell{c}{$0.028624$\\($0.294459$)}        &\tabincell{c}{$0.22002$\\($0.132597$)}         &\tabincell{c}{$0.84958$\\($0.023042$)}\\
  $  $         &BS-H    &\tabincell{c}{$0.078299$\\($0.286996$)}        &\tabincell{c}{$0.152854$\\($0.219707$)}         &\tabincell{c}{$0.948988$\\($0.095024$)}\\
\hline
   $ $         &IG-N     &\tabincell{c}{$0.017771$\\($0.216084$)}         &\tabincell{c}{$0.09206$\\($0.163681$)}        &\tabincell{c}{$0.38907$\\($0.052588$)}  \\
  CAN-USD &BS-I      &\tabincell{c}{$0.029684$\\($0.294301$)}         &\tabincell{c}{$0.22317$\\($0.132374$)}        &\tabincell{c}{$0.85063$\\($0.022957$)}      \\
  $  $        &BS-H     &\tabincell{c}{$0.075477$\\($0.286737$)}         &\tabincell{c}{$0.147718$\\($0.219266$)}        &\tabincell{c}{$0.933576$\\($0.094684$)}    \\
\bottomrule
\end{supertabular}
\end{center}
\vskip 2mm
\par 
Table 2 shows the relative pricing error (RPE) and NSE (in parenthesis) for Quanto options constructed using exchange rates between different currencies, including EUR-USD, UK-USD and CAN-USD. We choose options with different moneyness ($\frac{K_f}{X_t}\!<\!0.98$: in-the-money (ITM), $0.98\!<\!\frac{K_f}{X_t}\!<\!1.02$: at-the-money (ATM), $\frac{K_f}{X_t}\!>\!1.02$: out-of-the-money (OTM)) and compare the Bayesian method with IGN candidate densities with the BS-I and BS-H. We see that the adoption of three exchange rate processes leads to similar pricing errors, and the RPE increases with the increase of strike prices. Moreover, the Bayeisan method established in the paper always performs best comparing with the BS-I and BS-H models. Therefore, our paper provides an alternative estimation method of Quanto option prices  with investors from different countries to invest foreign options on the same underlying asset. 

We choose various sample sizes $T=140, 740, 1340$ and $1840$ days to make Bayesian inference on unknown parameters, and further test if the final pricing results are sensitive to the changes of sample size. 
 \begin{figure}[H]
  \centering
  \subfigure[Pricing results under the IGN candidate density with different underlying sample sizes]{
  \includegraphics[height=7cm,width=15cm]{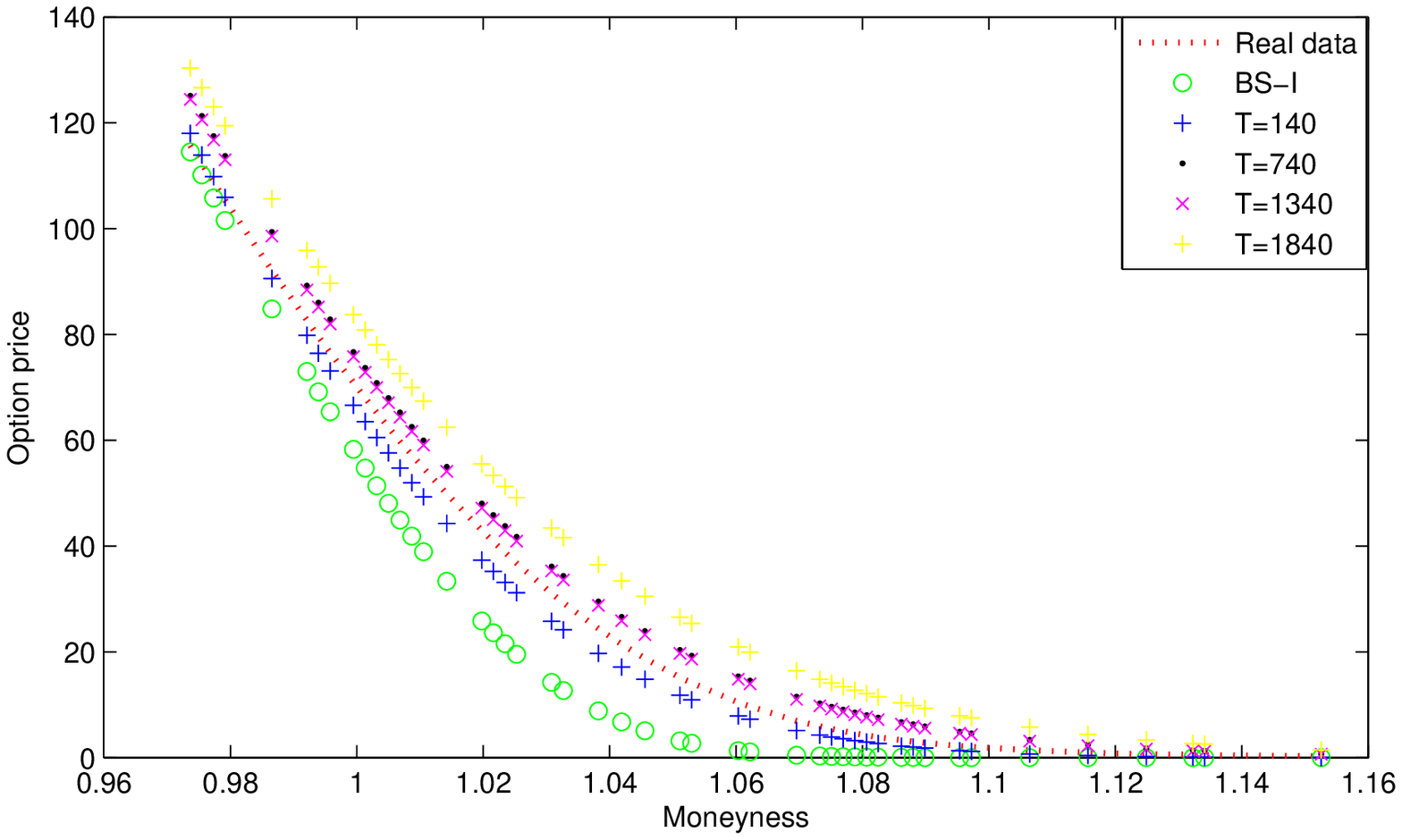}}
    \end{figure}
   \addtocounter{figure}{0} 
\begin{figure}[H]
\centering
  \subfigure[Pricing results of MNC with different underlying sample sizes]{
  \includegraphics[height=7cm,width=15cm]{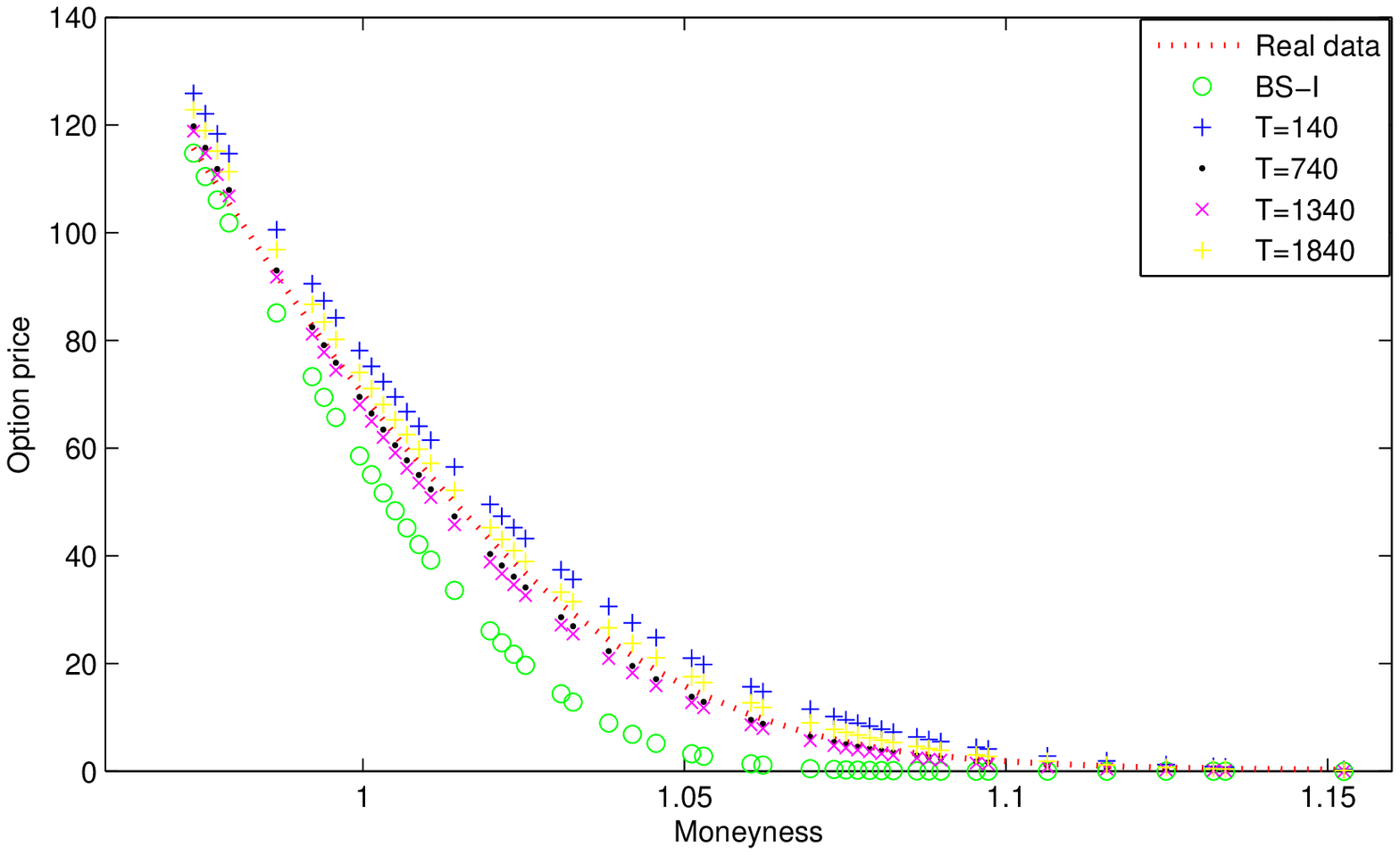}}
  \caption{Pricing results with different underlying sample sizes}
\end{figure}

Fig.\,5 plots the pricing performance under different underlying sample sizes by comparing with real market data and BS-I model prices. As we can see from Fig.\,5 (a), our method with IGN candidate densities is robust with different sample sizes, and it can achieve the best pricing performance when $T=140$ days. This conclusion is also valid for the Bayesian method with TTN and TNN candidate densities. However, Fig.\,5(b) shows that the Bayesian method with conjugate priors presents the best pricing result when $T=1840$ days. Therefore, we obtain the similar conclusion as in Ref.\,\cite{Rombouts(2014)}, that is, the Bayesian methods constructed in this paper could make positive contribution to pricing options on underlying assets for which less sample data is available. This advantage is extremely meaningful in some cases of the limited disclosure information for those newly listed firms, where the Bayesian method has potentials to do estimations and predictions.

\section{Conclusion}
This paper is concerned with the pricing of Quanto options using Bayesian methods with the randomness of parameters being considered. We performed Bayesian inference on the estimation of volatilities of foreign asset and exchange rate and the correlation between them. Then we considered four different types of Quanto options and explained how to compute option prices based on the Bayesian prediction technique as a by-product of Bayesian posterior density. 

To the best of our knowledge, Bayesian method has been widely used in the pricing of one-asset options, but less efforts has been made to price options with more than one asset.  Hence this paper is an extension work and serves as a basis for the application of Bayesian methods in the pricing of options with more unknown parameters. 
\par
In the empirical studies, the Metropolis with-in Gibbs algorithm with different candidate densities is presented for posterior sampling. Real Quanto options data with different maturities and strike prices are constructed in a rigorous way. The posterior predictive distribution is made use of as a goodness-of-fit testing to compute option prices. By comparing with the maximum likelihood estimation results, the Bayesian estimation results with informative conjugate priors and the real market prices, we found that the Bayesian method established in this study for Quanto options pricing has obvious advantages for all moneyness categories and has more potentials when less data for underlying assets is available. The MLE and Bayesian methods with conjugate priors usually overestimate and underestimate market option data, respectively. In addition, by constructing Quanto options with exchange rates between different currencies and testing the pricing results, we found that our method provides an alternative estimation method of Quanto option prices with investors from different countries to invest foreign options on the same underlying asset.
\par
Although we concentrated on the pricing of Quanto options throughout the paper, the framework established in this paper also applies to price other types of options with two assets, including exchange options, better-of options etc.\,.
In addition, we can find here, it will be an interesting future research topic to further extend the framework to more general stochastic processes, such as mean-reverting process, jump-diffusion process and the pure jump L$\mathrm{\acute{e}}$vy process.

\section*{Acknowledgements}
This work was supported by National Natural Science Foundation of China (7117 1077(Y.L.)), the Canada Research Chair Program (230720(J.W.)), the Natural Sciences and Engineering Research Council of Canada (105588-2011(J.W.)). The first author greatly acknowledges the financial support from China Scholarship Council.

\section*{References}
\small
\bibliographystyle{plain}
\bibliography{Quanto}

\end{document}